\documentclass[12pt]{article}
\usepackage{a4wide}
\usepackage{bbold}
\usepackage{amsmath,amssymb,bbm}
\usepackage{multirow}
\usepackage{comment}
\newsavebox{\uuunit}
\sbox{\uuunit}
    {\setlength{\unitlength}{0.825em}
     \begin{picture}(0.6,0.7)
        \thinlines
        \put(0,0){\line(1,0){0.5}}
        \put(0.15,0){\line(0,1){0.7}}
        \put(0.35,0){\line(0,1){0.8}}
       \multiput(0.3,0.8)(-0.04,-0.02){12}{\rule{0.5pt}{0.5pt}}
     \end {picture}}
\newcommand {\unity}{\mathord{\!\usebox{\uuunit}}}

\def\2{\frac12}
\def\4{\frac14}

\newcommand{\be}{\begin{equation}}
\newcommand{\ee}{\end{equation}}
\newcommand{\bea}{\begin{eqnarray}}
\newcommand{\eea}{\end{eqnarray}}
\def\a{\alpha}

\begin{document}

\begin{titlepage}%1
\begin{center}

\hfill UG-11-01 \\ \hfill KCL-MTH-11-01

\vskip 1.5cm

{\Large \bf  String Solitons and T-duality}

\vskip 1cm

{\bf Eric A.~Bergshoeff\,$^1$ and Fabio Riccioni\,$^2$}

\vskip 25pt

{\em $^1$ \hskip -.1truecm Centre for Theoretical Physics,
University of Groningen, \\ Nijenborgh 4, 9747 AG Groningen, The
Netherlands \vskip 5pt }

{email: {\tt E.A.Bergshoeff@rug.nl}} \\

\vskip 15pt

{\em $^2$ \hskip -.1truecm Department of Mathematics, Kings College
London, \\ Strand London, WC2R 2LS UK \vskip 5pt }

{email: {\tt Fabio.Riccioni@kcl.ac.uk}} \\

\end{center}

\vskip 0.5cm

\begin{center} {\bf ABSTRACT}\\[3ex]
\end{center}

We construct for arbitrary dimensions a universal T-duality
covariant expression for the Wess-Zumino terms of supersymmetric
String Solitons in toroidally compactified  string theories with 32
supercharges. The worldvolume fields occurring in the effective
action of these String Solitons form either a vector or a tensor
multiplet with 16 supercharges. We determine the dimensions of the
conjugacy classes under T-duality to which these String Solitons
belong. We do this in two steps. First, we determine the T-duality
representations of the $p$-forms of maximal supergravities that
contain the potentials that couple to these String Solitons. We find
that these are $p$-forms, with $D-4\le p\le 6$ if $D \ge 6$ and with
$D-4\le p\le D$ if $D < 6$, transforming in the antisymmetric
representation of rank $m=p+4-D\le 4$ of the T-duality symmetry
$\text{SO}(10-D,10-D)$. All branes support vector multiplets except
when $m=10-D$. In that case the T-duality representation splits, for
$D<10$, into a selfdual and anti-selfdual part, corresponding to
5-branes supporting either a vector or a tensor multiplet. In a
second step we show that only certain well-defined lightlike
directions in the anti-symmetric tensor representations of the
T-duality group correspond to supersymmetric String Solitons. These
lightlike directions define the  conjugacy classes.
%Eric
As a by-product we show how by a straightforward procedure all
solitonic fields of maximal supergravity are derived using the
Kac-Moody algebra $\text{E}_{11}$.

\end{titlepage}

\newpage
\setcounter{page}{1} \tableofcontents

\newpage

%%%%%%%%%%%%%%%%%%%%%%%%%%%%%%%%%%%% %%%%%%%%%%%%%%%%%%%%%%%%%%%%%%%%%%%%%
%%%%%%%%%%%%%%%%%%%%
\setcounter{page}{1} \numberwithin{equation}{section}

\section{Introduction}
Supergravity theories provide important information about the branes
of string theory, of which they are the low-energy limit. The
(bosonic) physical states of IIA and IIB supergravity are described
by the metric and a set of $p$-form potentials, with $0\le p \le 4$.
These latter fields couple to (half-supersymmetric) electric
$(p-1)$-branes whereas their Poincare duals couple to
(half-supersymmetric) magnetic $(7-p)$-branes.\,\footnote{This
excludes the IIA/IIB dilaton, but includes the axion of IIB
supergravity which couples to the D-instanton, or ``D(-1)-brane'',
whereas its dual RR 8-form potential couples to the D7-brane. The
IIA/IIB dilatonic dual 8-form potential will be discussed in this
paper.} It turns out that IIA and IIB supergravity can be extended
with 9-form and 10-form potentials that do not describe any physical
degrees of freedom \cite{Bergshoeff:1996ui,Bergshoeff:2005ac}.
Nevertheless, the full (on-shell) supersymmetry algebra can be
realised on these fields. It is therefore perfectly legitimate to
add them to the supergravity theory and to see whether they couple
to branes as well. Perhaps the best known example of such a field is
the 9-form potential of IIA supergravity \cite{Bergshoeff:1996ui}.
Its equation of motion follows from a duality relation between the
10-form curvature of this potential and the Romans mass parameter
$m$ \cite{Romans:1985tz}. This potential couples to the D8-brane, or
domain-wall, of Type IIA string theory. One can also add a set of
10-form potentials to IIA/IIB supergravity \cite{Bergshoeff:2005ac}.
In the case of IIB supergravity one of these fields couples to the
D9-brane. But there are more 10-form potentials on which  the
IIA/IIB supersymmetry algebra can be realised. Whether these other
potentials couple to branes as well is less clear. Recently, the
U-duality representations of the $(D-1)$-form and $D$-form
potentials of maximal supergravity in $D<10$ dimensions have been
determined by using the embedding tensor technique \cite{deWit:2008ta}.
Independently, they have been determined
\cite{Riccioni:2007au,Bergshoeff:2007qi} by making use of the
properties of the very extended Kac-Moody algebra $\text{E}_{11}$
\cite{West:2001as}.  The number of these potentials
becomes quite large in lower dimensions.

In this paper we wish to address the question of which $p$-form
potentials of maximal supergravities  correspond to supersymmetric
branes in (compactified) string theory. In doing so it is important
to first specify which kind of branes we wish to consider. We will
only consider branes for which we can construct a gauge-invariant
Wess-Zumino (WZ) term describing the coupling of the supergravity
potentials to the brane. The construction of such a gauge-invariant
WZ term in itself is straightforward. All one needs is the
transformation rules of the potentials under the different gauge
symmetries where for every (pull-back of the) gauge parameters one
introduces a corresponding worldvolume potential. We will impose the
non-trivial requirement that these worldvolume potentials fit into a
supermultiplet with 16 supercharges. The branes satisfying this
condition have the necessary ingredients to make the construction of
a kappa-symmetric worldvolume action possible. It is therefore
suggestive that these branes are well-defined in string
theory.\footnote{For branes with co-dimension 2 one needs to
integrate over the moduli space to obtain finite-energy
configurations. Branes with co-dimension 0, i.e.~space-time filling
branes, need to be combined with orientifold planes to define string
theories with 16 supercharges. In this paper we will not consider
these issues but only consider single branes by themselves.} We
stress that  this does {\sl not} imply that the remaining  $p$-form
potentials have nothing to do with branes at all. Their mere
existence as part of maximal supergravity is suggestive and perhaps,
when a proper non-perturbative formulation of string theory has been
developed, we will understand how these potentials couple to some
class of branes.

It is insightful to classify the branes according to the way their
tension $T$ scales with the string coupling constant $g_s$ in the
string frame. This can be specified by an integer number $\alpha$
via
\begin{equation}
T\ \sim\ (g_s)^\alpha\,.
\end{equation}
It turns out that in string theory $\alpha \le 0$. The highest
values of $\alpha$ correspond to the following kind of branes:
\begin{equation}
\alpha=0: \text{Fundamental Branes};\hskip .5truecm \alpha=-1:
\text{D-branes};\hskip .5truecm \alpha=-2: \text{Solitonic
Branes}\,.
\end{equation}
Our understanding of the branes with $\alpha =-3,-4,\cdots$ is
rather limited. In the lower dimensional theories, which result from
dimensional reduction from eleven dimensions, we may write
$g_s=<e^\phi>$ where the dilaton $\phi$ refers to any of the
compactified directions. Selecting this dilaton direction
corresponds to decomposing the $D$-dimensional U-duality group with
respect to T-duality as
  \begin{equation}
  \text{E}_{11-D(11-D)} \supset \text{SO}(10-D,10-D) \times \mathbb{R}^+
  \quad . \label{UdualityisTdualitytimesRplus}
  \end{equation}
%ericrevision%
The dependence of the brane tension on this dilaton, i.e.~the number
$\alpha$, is determined by the rank $p$ of the
corresponding $p$-form potential and the weight $w$ with which this
potential transforms under the $\mathbb{R}^+$-symmetry
\cite{Bergshoeff:2010xc}.\footnote{The number $\alpha$ has a natural group theory interpretation in terms of the
Kac-Moody algebra $\text{E}_{11}$ \cite{Cook:2008bi} (see also
\cite{Cook:2009ri}). We assume here that the brane
tension contains a leading term which only depends on the dilaton.
In general, this need not be the case.  To determine the full
dependence of $T$ on all fields we need the supersymmetry rules of
the $p$-form potentials. We will not investigate this further in
this work.} This conserved number is thus defined for any
supergravity field, regardless of whether it corresponds to a brane
or not.
%%Eric2%%
%We will show at the end of this paper that the very-extended Kac-Moody algebra  $\text{E}_{11}$ can be used to conveniently
%pick out all potentials of a given supergravity theory that
%correspond to a fixed value of $\alpha$.

%ericrevision%
In this work we will consider the Wess-Zumino terms of branes in $D\le 10$ dimensions for a fixed value of $\alpha$, i.e.~according to
T-duality representations. This is to be distinguished from the vast literature on S-duals of brane actions (including the kinetic terms)
in $D=10$  dimensions, see, e.g., \cite{earlier}.
In our  previous work we only considered the Fundamental Branes and
D-branes of (compactified) string theory \cite{Bergshoeff:2010xc}.
Since the Fundamental Branes have the highest value of $\alpha$,
only Fundamental Branes themselves can end on them. In practice, we
find that only Fundamental 0-Branes, i.e.~wrapped strings, may end
on the Fundamental String.  The construction of a gauge-invariant WZ
term therefore only requires world-volume scalars that fit into a
scalar multiplet. In $D=10$ dimensions only embedding scalars are
needed and the WZ term is given by the well-known expression
\begin{equation}
  {\cal L}^{\text{D=10}}_{\text{WZ}}\text{(Fundamental String)} = B_2\,,
  \end{equation}
where $B_2$ is the (pull-back of the) NS-NS 2-form. In $D < 10$
wrapped strings can end on the Fundamental String and the
corresponding WZ term gets accordingly modified with extra
world-volume scalars \cite{Hull:2004in}:
\begin{equation}\label{WZstring}
  {\cal L}^{\text{D} < \text{10}}_{\text{WZ}}\text{(Fundamental String)} =  B_2 + \eta^{AB} {\cal F}_{1, A} B_{1 ,B}\,,
  \end{equation}
where $B_{1,A}$ are the NS-NS 1-forms and  ${\cal F}_{1,A}$ are the
1-form world-volume curvatures of the extra scalars. Both  transform
as a vector, indicated by the index $A$, under the T-duality group
$\text{SO}(d,d)$ with $d=10-D$. The number of extra scalars is twice
the number of compactified dimensions in line with doubled geometry
\cite{Hull:2004in}. The WZ term for the Fundamental 0-branes
themselves does not contain these extra scalars and is given by
(omitting the explicit vector-index $A$)
  \begin{equation}\label{WZ0branes}
  {\cal L}^{\text{D} < \text{10}}_{\text{WZ}}\text{(Fundamental
  0-Branes)} = B_1\,.
  \end{equation}

We next consider the D-branes. Since they have the one-to-highest
value of $\alpha$, only Fundamental Branes can end on them.\,\footnote{
Here and in the rest of the paper we only consider the objects that are electrically charged with respect
to the worldvolume fields. For each such object there is a corresponding magnetic object that can end on
the same brane. For instance, in the case of a D$p$-brane, the magnetic version of the fundamental string is
a D$(p - 2)$-brane and the magnetic version of a fundamental particle is a D$(p - 1)$-brane.}
 In
$D=10$ dimensions there are only Fundamental Strings that can end on
D-branes and, accordingly, the WZ term gets deformed by an extra
Born-Infeld worldvolume vector, with 2-form curvature ${\cal F}_2$:
\begin{equation}\label{WZterm}
  {\cal L}^{\text{D=10}}_{\text{WZ}}\text{(D-branes)} = e^{{\cal F}_2}C\,.
  \end{equation}
Here  $C$ stands for the formal sum of all RR potentials. In
\cite{Bergshoeff:2010xc} we derived the T-duality-covariant
expression of the D-brane WZ terms in $D < 10$ dimensions. Since now
both wrapped and un-wrapped Fundamental Strings can end on the
D-branes we get a further deformation by the extra worldvolume
scalars \cite{Bergshoeff:2010xc}:
 \begin{equation}\label{WZterm2}
  {\cal L}^{\text{D} < \text{10}}_{\text{WZ}}\text{(D-branes)} = e^{{\cal F}_2}e^{{\cal F}_{1,A}\Gamma^A}C\,,
  \end{equation}
where $\Gamma^A$ are the gamma-matrices of $\text{SO}(d,d)$. The
reason for the existence of the general expression \eqref{WZterm2}
is that in any dimension the fundamental potentials transform as a
singlet (2-form) and vector (1-form) under T-duality while the
D-brane potentials ($p$-forms) transform as (chiral) spinor
representations of the same duality group.

In this paper we wish to continue the analysis of
\cite{Bergshoeff:2010xc} and consider the next set of branes,
i.e.~the String Solitons. The analysis  becomes now more subtle due
to two reasons. First of all, both Fundamental Branes as well as
D-branes may end on String Solitons. This leads to many world-volume
$p$-form potentials which must fit into one of the two available
world-volume supermultiplets with 16 supercharges: the vector
multiplet in ten dimensions or less (with 1 vector) or the
six-dimensional tensor multiplet (with 1 self-dual 2-form
potential). In general, we will obtain too many world-volume
potentials to fit any of these two multiplets. There are two ways to
lower the number of the world-volume  potentials. One way is to
impose world-volume duality relations. Another way is by making
suitable redefinitions of the target space solitonic,
i.e.~$\alpha=-2$, potentials with terms that are quadratic in the
$\alpha=-1$ RR potentials. In many cases this lowering of the number
of world-volume potentials turns out not to be enough. Unlike the
Fundamental Branes and the D-branes  we find that many of the
solitonic potentials of maximal supergravity do not satisfy our
criterion that they must couple to a String Soliton via a
gauge-invariant WZ term that only contains world-volume potentials
that fit into a vector or a tensor multiplet. The fact that we find
branes that satisfy our criterion is possible due to a beautiful and
intricate interplay between target space and worldvolume
supersymmetry and electromagnetic duality.

To classify the cases in which supersymmetric String Solitons exist
we proceed in two steps. First, we determine the T-duality
representations of the potentials that contain the ones that couple
to supersymmetric String Solitons. We find that these are
antisymmetric tensor representations. Next, we determine the
conjugacy classes within the T-duality representations to which the
supersymmetric String Solitons belong. We will show that these
conjugacy classes are defined by a specific set of lightlike
directions within the antisymmetric tensor representation. The
phenomenon that branes only form a conjugacy class within a given
T-duality representation does not occur for Fundamental Branes and
D-branes simply because in these cases the T-duality representations
involved do not contain non-trivial conjugacy classes. It has
however a well-known analogue in the case of ten-dimensional
7-branes with respect to the S-duality group
$\text{SL}(2,\mathbb{R})$. The 8-form potentials of IIB supergravity
that contain the potential that couples to the D7-brane are in the
${\bf 3}$ representation of
$\text{SL}(2,\mathbb{R})$.\,\footnote{The IIB supersymmetry algebra
closes on a triplet of 8-forms provided that a duality relation
with the two scalars of IIB supergravity holds \cite{modlt}. This relation implies that one
combination of the curvatures of the three 8-form
potentials (multiplied by scalar-dependent factors) vanishes, in agreement with the fact that the theory
contains only two scalars. This relation does not play a role in the
present discussion.} It turns out that the D7-brane and its S-dual
belong to a 2-dimensional conjugacy class within this triplet. The
remaining component of the triplet does not correspond to a brane.
To see this it is convenient to use a real $\text{SO}(2,1)$ notation
since that resembles most the discussion in the case of T-duality.
The crucial point is that in the construction of the WZ term for the
triplet of 7-branes there is a coupling between the curvatures of
the worldvolume 1-forms and the target space 6-forms of the form
%%Eric2%%
\begin{equation}
{\cal L}_{\text{WZ}}^{\text{D=10}}(\text{7-branes}) \ \ \sim \ \
A_{8,i}\ +\ {\cal F}_2\Gamma_i A_6\ +\ \cdots\,,\hskip 1truecm i=1,2,3\,.
\end{equation}
Here ${\cal F}_2$ and $A_6$ transform as two-component spinors of
$\text{SO}(2,1)$, that contain the Born-Infeld vector and the RR
6-form together with their S-duals, respectively. We want the ${\cal
F}_2\Gamma_i A_6$ term to describe only a coupling between the
Born-Infeld vector and the RR 6-form or the S-dual version of this.
In particular, worldvolume supersymmetry requires that the
Born-Infeld vector and its S-dual do not occur at the same time.
Therefore, we want the $2\times 2$ gamma matrix $\Gamma_i$ to act as
a projection operator which picks out the correct  component of the
2-dimensional spinor ${\cal F}_2$.  To achieve this it is convenient
to introduce a lightcone basis $i=(+,-,3)$. One can easily convince
oneself that the lightlike directions $\Gamma_+$ and $\Gamma_-$ that
square to zero define the conjugacy class containing the D7-brane
and its S-dual, which is an $\alpha=-3$ object. We will confirm in
this paper that the remaining solitonic 8-form potential $A_{8,3}$,
which has weight $\alpha=-2$, does not couple to a supersymmetric
String Soliton.\,\footnote{Using $\text{SL}(2,\mathbb{R})$ notation
the conjugacy classes of $\text{SL}(2,\mathbb{R})$ are labelled by
$\det Q$ where $Q$ is the $2\times 2$  charge matrix. The D7-brane
and its S-dual belong to the $\det Q=0$ conjugacy class while the
remaining 8-form potential, the one that does not correspond to a
brane, belongs to a $\det Q <0$ conjugacy
class, see the last reference of \cite{earlier}. This is to be distinguished from the
7-branes at the three orbifold points in the
$\text{SL}(2,\mathbb{R})$ moduli space. One of these branes is the
D7-brane and belongs to the $\det Q=0$ conjugacy class while the
branes at the other 2 orbifold points belong to the $\det Q>0$
conjugacy class. The latter branes can be viewed as bound states of
the D7-brane with the S-dual of the D7-brane or anti-D7 brane, see
e.g.~\cite{Bergshoeff:2006jj}.} Our analysis of the conjugacy
classes with respect to T-duality is very similar to the S-duality
discussion above except that one is now dealing with the group
$\text{SO}(d,d)$ instead of $\text{SO}(2,1)$.

A second feature that makes the analysis of String Solitons more
subtle is that, whereas all Fundamental Branes and D-branes in
$D<10$ dimensions can be obtained via dimensional reduction of the
$D=10$ Fundamental String and D-branes, the same is not the case for
the String Solitons.
%%Eric2%%
One way to understand the 10-dimensional
origin of all lower-dimensional T-duality multiplets of solitonic
potentials is to add to the standard supergravity fields extra
solitonic fields of mixed symmetry without upsetting the counting of degrees of freedom.
 At the end of this paper we will show that the ten-dimensional
mixed-symmetry solitonic fields that are needed to generate all the
solitonic fields in lower dimensions are precisely the ones predicted by the very
extended Kac-Moody algebra $\text{E}_{11}$. It is not understood how
to realise the full IIA/IIB supersymmetry algebra on these mixed
symmetry potentials. This can only be achieved for linearised
supersymmetry. However, these mixed-symmetry fields contain
important information in the sense that, after reduction, they give
rise to standard $p$-form potentials on which the non-linear
supersymmetry algebra can be realised. In this paper we will derive
a useful procedure of how to collect {\sl all} (i.e.~both $p$-form
and mixed-symmetry) fields, predicted by $\text{E}_{11}$, for a
given value of $\alpha$. We will find that for $\alpha$ odd the
representations of the IIA and IIB fields differ but that for
$\alpha$ even they coincide. This applies in particular to the
fundamental and solitonic potentials.

It is instructive to see how in $D=10$ dimensions the construction
of a gauge-invariant WZ term can be achieved with a supermultiplet
of world-volume potentials. Both Type IIA and Type IIB string theory
contain a single solitonic 6-form potential $D_6$, which couples to
the NS5A-brane and NS5B-brane, respectively. The corresponding
gauge-invariant WZ terms are given by (see section 3)
\begin{eqnarray}\label{WZNSA}
{\cal L}^{\text{D=10}}_{\text{WZ}}(\text{NS5A-brane}) &=& D_6 +
{\cal G}_5C_1-{\cal G}_3C_3 +{\cal
G}_1C_5\,,\\[.2truecm]
{\cal L}^{\text{D=10}}_{\text{WZ}}(\text{NS5B-brane}) &=& D_6 +
{\cal G}_4C_2-  {\cal G}_2C_4\,,\label{WZNSB}
\end{eqnarray}
where $C_p$ is a target space RR $p$-form potential and where we
have introduced world-volume $p$-form potentials $c_p$ with
curvatures ${\cal G}_{p+1}$. These expressions can be
straightforwardly obtained by applying a Noether procedure,
replacing each parameter in the  transformation rule of $D_6$  by a
corresponding world-volume potential. The NS5A-brane contains the
worldvolume potentials $c_0, c_2$ and $c_4$. Imposing a duality
relation between them we are left with a scalar and a self-dual
2-form. Together with the 4 embedding scalars these form the bosonic
sector of an 8+8 tensor multiplet in the six-dimensional
worldvolume. The extra $c_0$ scalar indicates that this solitonic
5-brane has an 11-dimensional origin as the M5-brane. On the other
hand, the NS5B-brane contains the worldvolume potentials $c_1$ and
$c_3$. Imposing a duality relation between them we are left with one
vector. Together with the 4 embedding scalars these form the bosonic
sector of an 8+8 vector multiplet in the six-dimensional
worldvolume.

It turns out that in $D=10$ there are three more solitonic $p$-form
potentials $D_8\,, D_{10}$ and $D_{10}^\prime$, see Table
\ref{IIAIIBD}. However, unlike $D_6$, these potentials cannot be
associated with a supersymmetric String Soliton. These findings are
consistent with our earlier analysis of Supersymmetric String
%%Eric2%%
Solitons, see
\cite{Bergshoeff:2005ac,Bergshoeff:2006ic} and the last reference of \cite{earlier}. The
criterion we imposed in these references was
that a cancellation under supersymmetry was taking place between the
Nambu-Goto kinetic term and the leading solitonic potential in the
WZ term. This requirement leads to an expression for the brane
tension and the BPS condition. The conclusion  based upon this
criterion  was
that no Supersymmetric Soliton could be associated with $D_8\,,
D_{10}$ and $D_{10}^\prime$. In this work we reach the same
conclusion based upon the independent criterion that a
gauge-invariant WZ term must exist with worldvolume potentials that
fit into a supersymmetry multiplet.

We find that in $D<10$ dimensions the above results generalise in a
T-duality covariant way as follows. In $D$ dimensions the
$T$-duality representations of the potentials of maximal
supergravity that contain the potentials that couple to
supersymmetric String Solitons are anti-symmetric tensor
representations of the T-duality group.
 More precisely, we find that these potentials are
$p$-forms, with $D-4\le p\le 6$ if $D \ge 6$ and with $D-4\le p\le
D$ if $D < 6$, transforming in the antisymmetric representation of
rank $m=p+4-D\le 4$ of the T-duality symmetry
$\text{SO}(10-D,10-D)$. We have summarised this result in Table
\ref{result}. As discussed above the Supersymmetric Solitons form
conjugacy classes within these T-duality representations. These
conjugacy classes are described in Table \ref{conjugacyclasses}. The
10-dimensional origin of these conjugacy classes comes from $D_6$,
together with a number of mixed-symmetry fields. The fact that these
conjugacy classes contain among its components reductions of the
known supersymmetric NS5A and NS5B branes of string theory
guarantees that all other components of the same conjugacy class
correspond to Supersymmetric String Solitons as well. All the
remaining T-duality multiplets that do not contain potentials that
can be associated with supersymmetric String Solitons, see Table
\ref{NSformsanyD}, come from either $D_8\,, D_{10}$ and
$D_{10}^\prime$, together with a number of the mixed symmetry
fields, or from  mixed symmetry fields only.

\begin{table}[t]
\begin{center}
\begin{tabular}{|c|c|c|c|}
\hline
$p$&solitonic potential&vector/tensor&vector\\[.1truecm]
\hline \rule[-1mm]{0mm}{6mm} $D-4$ & $D_{D-4}$&IIB/IIA& $4 \le D < 10$\\[.05truecm]
\hline \rule[-1mm]{0mm}{6mm} $D-3$ & $D_{D-3, A}$ &$D=9$ & $3 \le D < 9$\\[.05truecm]
\hline \rule[-1mm]{0mm}{6mm} $D-2$ & $D_{D-2, AB}$& $D=8$& $2 \le D < 8$\\[.05truecm]
\hline \rule[-1mm]{0mm}{6mm} $D-1$ & $D_{D-1, ABC}$&$D=7$& $1 \le D < 7$\\[.05truecm]
\hline \rule[-1mm]{0mm}{6mm} $D$ & $D_{D, ABCD}$&$D=6$& $0 \le D < 6$\\[.05truecm]
\hline
\end{tabular}
\end{center}
\caption{\sl The T-duality multiplets of solitonic potentials that
contain the ones that couple to supersymmetric String Solitons. The
indices $A,B,...$ are vector indices of $\text{SO}(d,d)$, and they
are always meant to be antisymmetrised. The last two columns
indicate for which dimensions the worldvolume dynamics is described
by a vector/tensor multiplet or by vector multiplets only, see the
text. }
  \label{result}
\end{table}

Only 5-brane String Solitons can have both vector and tensor
world-volume multiplets. The corresponding 6-form potentials have
the peculiarity of always transforming as the reducible
representation of $\text{SO}(d, d)$ with $d$ antisymmetric indices,
which splits in self-dual and anti self-dual irreducible
representations. We will see that the self-dual representations
correspond to String Solitons with vector multiplets, like the
NS5B-brane, whereas the anti-selfdual representations correspond to
String Solitons with tensor multiplets, like the NS5A-brane:
\begin{equation}
D^+_{6,A_1\cdots A_d}\,:\ \text{vector multiplets}\,,\hskip 2truecm
D^-_{6,A_1\cdots A_d}\,:\ \text{tensor multiplets}\,.
\end{equation}
The $D=10$ case is special in the sense that the two 6-form
potentials belong to two different theories (IIA and IIB), see Table
\ref{result}.

In all cases we find that the gauge-invariant WZ term of the
%eric
Supersymmetric Solitons is given by (omitting the anti-symmetric
$A$-indices)
\begin{equation}
 {\cal L}^{\text{D}\le 10}_{\text{WZ}}(\text{susy soliton})
 = e^{{\cal F}_1}\left(D +{\overline {\cal G}}\Gamma C\right)\,,\label{WZD<10}
\end{equation}
where $C$ and $D$ represent the RR and solitonic target space
potentials, respectively, while ${\cal F}_1$ and ${\cal G}$ are the
worldvolume curvatures of the fundamental worldvolume scalars and RR
worldvolume potentials, respectively. In this expression, whose
precise form (containing all indices and signs) can be found in
eq.~\eqref{WZtermuniversalexpression}, it is understood that the
String Solitons form a  conjugacy class within the full T-duality
representation. This conjugacy class is defined by specifying a
certain set of lightlike directions, see eq.~\eqref{informula}.
Formula \eqref{WZD<10}, together with a description of the conjugacy
classes, constitutes the main result of this paper. This formula
includes the ten-dimensional expressions \eqref{WZNSA} and
\eqref{WZNSB} as special cases with the understanding that in the
IIA (IIB) case we only consider the even-form (odd-form)
world-volume potentials. The special feature of the WZ-term
\eqref{WZD<10}, which is only possible for the antisymmetric
T-duality representations of Table \ref{result}, is that the
gauge-invariance of the WZ-term can be achieved {\sl without}
introducing a  Born-Infeld vector $b_1$ on the world-volume, i.e.~no
fundamental string can end on these supersymmetric String Solitons.
For all the other T-duality multiplets, see Table \ref{NSformsanyD},
one must introduce this Born-Infeld vector and this leads to too
many worldvolume potentials that cannot fit into a vector or tensor
multiplet.

This paper is organised as follows. In section 2 we derive in any
dimension the solitonic potentials by decomposing the U-duality
representations of all the potentials of any maximal supergravity
theory with respect to T-duality. The result is summarised in Table
\ref{NSformsanyD}. In section 3 we review and discuss the properties
of the ten-dimensional String Solitons. In particular, we show why
the solitonic  potentials  $D_8\,, D_{10}$ and $D_{10}^\prime$ do
not couple to supersymmetric String Solitons. Next, in section 4, we
derive the gauge transformations of all solitonic potentials in
$D<10$ dimensions. This will be used in section 5 to derive the
different gauge-invariant WZ terms of all $D<10$ String Solitons.
The final result is given in eq.~\eqref{WZtermuniversalexpression}.
In the same section we define the lightlike directions that define
the conjugacy class to which the supersymmetric String Solitons
belong, see eq.~\eqref{informula}. In section 6 we show that our
results agree with the classification of central charges of the
supersymmetry algebra in any dimension. In section 7 we discuss
how a possible ten-dimensional origin of the $D<10$ solitons leads one
to consider mixed-symmetry fields in ten dimensions hinting at an underlying $\text{E}_{11}$ structure.
We explain a procedure of how to extract out of $\text{E}_{11}$ all  fields of
IIA and IIB, both $p$-forms and mixed symmetry fields, for a given
value of $\alpha$. As a side-result we will show that for IIA/IIB
these fields coincide for even $\alpha$ but are different for odd
$\alpha$. Finally, in section 8 we present our conclusions and
discuss different consequences of our work. We also include an
Appendix, in which we discuss the properties of the spinors of
$\text{SO}(d,d)$ that are used in sections 4 and 5.

\section{Solitonic fields in $D<10$  dimensions}

In recent years it has been established that maximal supergravity theories do not only contain
$p$-form potentials, together with their duals, that describe physical degrees of freedom but also $(D-1)$-form
and $D$-form potentials that do not describe any physical degree of freedom. The U-duality representations of
these $p$-form potentials have been established by different means, all leading to the same answer.
In $D=10$ dimensions all $9$-form and $10$-form potentials that are consistent with the closure of the IIA or IIB supersymmetry algebra have been determined \cite{Bergshoeff:2005ac}. Upon toroidal reduction this leads to similar $(D-1)$-form and $D$-form potentials in $D < 10$ dimensions.
Their U-duality representations have been independently determined by a classification of gauged maximal supergravities
\cite{deWit:2008ta} as well as by making use of $\text{E}_{11}$  techniques \cite{Riccioni:2007au,Bergshoeff:2007qi}.

In this section we will derive the T-duality representations of the solitonic potentials in any
dimensions by decomposing these U-duality representations with
respect to T-duality according to eq.~\eqref{UdualityisTdualitytimesRplus} and by picking out the solitonic fields. In any dimension $\alpha$,
the number that determines how the brane tension scales with the string coupling constant,
is related to the $\mathbb{R}^+$ weight in the decomposition \eqref{UdualityisTdualitytimesRplus}
and to the rank of the form. We
will explicitly perform our analysis in dimensions higher than four,
but as we will see in section 7 our results are  general
and apply to  four and three dimensions as well. The resulting T-duality representations
take on a universal form for any dimension which we have
summarised in Table \ref{NSformsanyD}. Here and in the rest of the
paper we denote the solitonic fields with $D$, simply because the
$\alpha=0$ fields and the $\alpha=-1$ fields are denoted in the
literature and in this paper with $B$ and $C$ respectively. The
solitonic fields should not be confused with the dimension of
spacetime, which is also denoted with $D$. Actually, in
\cite{Bergshoeff:2010xc} the decomposition of the U-duality
representations with respect to the T-duality group $\text{SO}(d,d)$
was already performed, and the D-brane fields with $\alpha=-1$ were listed.
Here we have refined that analysis. We have listed the result  in Tables
\ref{qD=9}-\ref{qD=5} where we have given the value of $\alpha$ corresponding to
all T-duality representations in dimensions $5 \le D \le 9$. In these tables $w$ denotes the
$\mathbb{R}^+$ weight, and the U-duality group decomposes as in eq.
\eqref{UdualityisTdualitytimesRplus}. The conventions for the
representations and the $w$ weights are those of
\cite{Bergshoeff:2010xc}, which, in turn, are taken from
\cite{Slansky}.

\begin{table}[t]
\begin{center}
\begin{tabular}{|c|c|}
\hline \rule[-1mm]{0mm}{6mm} $(D-4)$-form & $D_{D-4}$ \\
\hline \rule[-1mm]{0mm}{6mm} $(D-3)$-form & $D_{D-3, A}$ \\
\hline \rule[-1mm]{0mm}{6mm} $(D-2)$-form & $D_{D-2} + D_{D-2, AB}$ \\
\hline \rule[-1mm]{0mm}{6mm} $(D-1)$-form & $D_{D-1, A} + D_{D-1, ABC}$\\
\hline \rule[-1mm]{0mm}{6mm} $D$-form & $D_D + D^\prime_D+ D_{D, AB}
+ D_{D, ABCD}$
\\
 \hline
\end{tabular}
\end{center}
  \caption{\sl Forms with $\a=-2$ in any dimension $D\geq 3$ (In three
  dimensions the first line is absent because the corresponding field does not exist). The indices $A,B,...$
  are vector indices of $\text{SO}(d,d)$, and they are always meant to be antisymmetrised. }
  \label{NSformsanyD}
\end{table}

\begin{table}[t]
\begin{center}
\begin{tabular}{|c|c||c|c|c|c|c|}
\hline \rule[-1mm]{0mm}{6mm} field & U repr & $\a = 0$ & $\a =-1$ & $\a= -2$ & $\a= -3$ & $\a=-4$ \\
\hline \hline \rule[-1mm]{0mm}{6mm} 1-form & ${\bf 2}_0 $ & $(1,0)$
&
$(-1,0)$ &  & &   \\
\cline{2-7} \rule[-1mm]{0mm}{6mm}  & ${\bf 1}_1$ & $(0,1)$  &  & & & \\
\hline \rule[-1mm]{0mm}{6mm} 2-form & ${\bf 2}_1$ & $(1,1)$ &
$(-1,1)$ &
 & &  \\
\hline \rule[-1mm]{0mm}{6mm} 3-form & ${\bf 1}_1$ & & $(0,1)$ & & &\\
\hline \rule[-1mm]{0mm}{6mm} 4-form & ${\bf 1}_2 $ &
& $(0,2)$ & & & \\
\hline \rule[-1mm]{0mm}{6mm} 5-form & ${\bf 2}_2$ & & $(1,2)$ &  $(-1,2)$ & & \\
\hline \rule[-1mm]{0mm}{6mm} 6-form & ${\bf 2}_3$ & & $(1,3)$ & $(-1,3)$ &  & \\
\cline{2-7} \rule[-1mm]{0mm}{6mm}  & ${\bf 1}_2$ &  & & $(0,2)$ & & \\
\hline \rule[-1mm]{0mm}{6mm} 7-form & ${\bf 3}_3$ & & $(2,3)$ &$(0,3) $ & $ (-2,3)$&  \\
\cline{2-7} \rule[-1mm]{0mm}{6mm}  & ${\bf 1}_3$ & & & $(0,3)$ & & \\
\hline \rule[-1mm]{0mm}{6mm} 8-form & ${\bf 3}_4$ & & $(2,4)$ &
$(0,4) $&
 $(-2,4)$ & \\
\cline{2-7} \rule[-1mm]{0mm}{6mm}  & ${\bf 2}_3$ & & & $(1,3) $ & $(-1,3)$ & \\
\hline \rule[-1mm]{0mm}{6mm} 9-form & ${\bf 4}_4$ &  & $(3,4)$ &
$(1,4)$ &
$ (-1,4)$  &  $(-3,4)$ \\
\cline{2-7} \rule[-1mm]{0mm}{6mm}  & $2\times{\bf 2}_4$ & & & $2 \times (1,4) $ & $ 2\times (-1,4)$& \\
 \hline
\end{tabular}
\end{center}
  \caption{\sl The decomposition of the $n$-form potentials of $D=9$ maximal
supergravity. The U-duality is $\text{SL}(2, \mathbb{R})\times
\mathbb{R}^+$. We denote with $(w_1, w_2 )$ the weights associated
to $\mathbb{R}^+\times \mathbb{R}^+ $. The weight under T-duality is
$w_1 - w_2$. The value of $\alpha$ is given by $ \alpha =
\frac{1}{2} (w_1 + w_2 -n )$. \label{qD=9}}
\end{table}

We now proceed with an analysis of all the $\alpha=-2$ fields listed
in Tables \ref{qD=9}-\ref{qD=5} and summarized in Table \ref{NSformsanyD}.
In all cases, the form of lowest
rank is a $D-4$ form, that transforms as a singlet under T-duality.
This form is the magnetic dual of the Fundamental 2-form $B_2$, and
we denote it by $D_{D-4}$.

Increasing the rank by one unit, one can see from the tables that in
any dimension one has solitonic  $D-3$-forms transforming as
vectors under T-duality. These fields are the duals of the
Fundamental fields $B_{1,A}$, and we denote them by $D_{D-3,A}$. A
particular case is the nine-dimensional one, in which the T-duality group
is $\text{SO}(1,1)$ and the vector representation splits into its
selfdual and anti-selfdual part. This can be seen in Table
\ref{qD=9}, which shows that the U-duality representation of the
6-form is reducible and each of the two irreducible components
contain either the selfdual or the anti-selfdual representation when
decomposed under T-duality.

The T-duality representation of the solitonic $D-2$-forms is
reducible in all cases. The reader can derive in all cases by
looking at the tables that the fields are $D_{D-2}$ and
$D_{D-2,AB}$, transforming as a singlet and as an antisymmetric
tensor respectively under T-duality. In $D=9$ these are actually two
singlets, and each singlet arises from each of the two irreducible
U-duality representations of the 7-forms. In $D=8$ the singlet and
the selfdual part of $D_{6,AB}$ arise from the decomposition of the
 6-form in the $({\bf 8,1})$ U-duality representation, while the anti-selfdual part of $D_{6,AB}$ is
the  6-form in the $({\bf 1,3})$ U-duality representation. In dimensions lower than eight $D_{D-2}$
and $D_{D-2,AB}$ arise from the same U-duality representation.

Proceeding in the same way, one finds  that the  solitonic
$D-1$-forms are $D_{D-1,A}$ and $D_{D-1,ABC}$, that is a vector and
an antisymmetric tensor with three indices under T-duality. In seven dimensions the
field $D_{D-1,ABC}$ splits into a selfdual and an anti-selfdual part. Clearly, such a three-index antisymmetric tensor
 does not occur in nine dimensions due to the fact that the corresponding $\text{SO}(1,1)$ T-duality representation does
not exist.

Finally, one can see that in all cases the solitonic $D$-forms are
two singlets $D_{D}$ and $D_{D}^\prime$, an antisymmetric tensor
with two indices $D_{D,AB}$ and and antisymmetric tensor with four
indices $D_{D,ABCD}$. The latter field does not exist in nine
dimensions. To summarise, one can see from the tables that the
column corresponding to $\alpha=-2$ gives in all cases the general
result given in Table \ref{NSformsanyD}.

\begin{table}[t]
\begin{center}
\begin{tabular}{|c|c||c|c|c|c|c|}
\hline \rule[-1mm]{0mm}{6mm} field & U repr & $\a=0$ &  $\a=-1$ & $\a=-2$ & $\a=-3$ & $ \a= -4$ \\
\hline \hline \rule[-1mm]{0mm}{6mm} 1-form & $({\bf
{\overline{3}}},{\bf 2}) $ & $({\bf 2},{\bf 2})_1$ &
$({\bf 1}, {\bf 2})_{-2}   $ & &  &\\
\hline \rule[-1mm]{0mm}{6mm} 2-form & $({\bf 3},{\bf 1})$ & $({\bf
1},{\bf 1})_{2}$ & $({\bf 2},{\bf 1})_{-1} $ &
 & & \\
\hline \rule[-1mm]{0mm}{6mm} 3-form & $({\bf {1}},{\bf 2})$ &  & $({\bf 1},{\bf 2})_0$ & & & \\
\hline \rule[-1mm]{0mm}{6mm} 4-form & $({\bf \overline{3}},{\bf 1})$
& & $({\bf 2},{\bf 1})_1$ &  $({\bf 1},{\bf 1})_{-2}$ &  & \\
\hline \rule[-1mm]{0mm}{6mm} 5-form & $({\bf 3},{\bf 2})$ & & $({\bf
1},{\bf 2})_2$ &  $({\bf 2},{\bf 2})_{-1}$ & & \\
\hline \rule[-1mm]{0mm}{6mm} 6-form & $({\bf 8},{\bf 1})$ & & $({\bf
2},{\bf 1})_3$ & $({\bf 1},{\bf 1})_{0}+ ({\bf 3},{\bf 1})_{0}$ & $({\bf 2},{\bf 1})_{-3} $  & \\
\cline{2-7} \rule[-1mm]{0mm}{6mm}  & $({\bf 1},{\bf 3})$ &  & & $({\bf 1},{\bf 3})_{0}$& & \\
\hline \rule[-1mm]{0mm}{6mm} 7-form & $({\bf 6},{\bf 2})$ & & $({\bf
1},{\bf 2})_4$ & $({\bf 2},{\bf 2})_{1}$&  $ ({\bf 3},{\bf 2})_{-2}$ & \\
\cline{2-7} \rule[-1mm]{0mm}{6mm}  & $({\bf \overline{3}},{\bf 2})$
& & &  $({\bf 2},{\bf 2})_{1}$& $
({\bf 1},{\bf 2})_{-2}$& \\
\hline \rule[-1mm]{0mm}{6mm} 8-form & $({\bf 15},{\bf 1})$ & &
$({\bf 2},{\bf 1})_5$ & $({\bf 3},{\bf 1})_{2} +({\bf 1},{\bf
1})_{2}$ & $({\bf 4},{\bf 1})_{-1} + ({\bf 2},{\bf
1})_{-1}$ & $({\bf 3},{\bf 1})_{-4} $ \\
\cline{2-7} \rule[-1mm]{0mm}{6mm}  & $({\bf {3}},{\bf 3})$ & & &
$({\bf 1},{\bf 3})_{2}$ & $({\bf 2},{\bf 3})_{-1}$
& \\
\cline{2-7} \rule[-1mm]{0mm}{6mm}  & $2 \times({\bf {3}},{\bf 1})$ &
& & $2 \times
({\bf 1},{\bf 1})_{2} $ & $2 \times({\bf 2},{\bf 1})_{-1} $ & \\
 \hline
\end{tabular}
\end{center}
  \caption{\sl  The decomposition of the $n$-form potentials of $D=8$ maximal
supergravity. The U-duality symmetry is $\text{SL}(3,\mathbb{R})
\times \text{SL}(2,\mathbb{R})$ and the T-duality is
$\text{SL}(2,\mathbb{R}) \times \text{SL}(2,\mathbb{R})$, while the
subscript denotes the $\mathbb{R}^+$-- charge $w$, which is related
to $\alpha$ by the equation $\alpha = -\frac{1}{3}(n-w)$.
\label{qD=8}}
\end{table}

 \begin{table}[t]
\begin{center}
\begin{tabular}{|c|c||c|c|c|c|c|}
\hline \rule[-1mm]{0mm}{6mm} field & U repr & $\a=0$ & $\a=-1$ & $\a=-2$ & $\a=-3$ & $\a=-4$\\
\hline \hline \rule[-1mm]{0mm}{6mm} 1-form & ${\bf \overline{10}}$ &
${\bf {6}}_2$  &
${\bf \overline{4}}_{-3}$ & & &\\
\hline \rule[-1mm]{0mm}{6mm} 2-form & ${\bf 5}$ &  ${\bf 1}_{4}$ &
${\bf 4}_{-1}$ &
& &\\
\hline \rule[-1mm]{0mm}{6mm} 3-form & ${\bf \overline{5}}$ & & ${\bf \overline{4}}_{1}$ & ${\bf 1}_{-4}$& & \\
\hline \rule[-1mm]{0mm}{6mm} 4-form & ${\bf 10}$ & & ${\bf {4}}_{3}$ & ${\bf 6}_{-2}$ & & \\
\hline \rule[-1mm]{0mm}{6mm} 5-form & ${\bf 24}$ &  &${\bf \overline{4}}_5$ & ${\bf 15}_0+ {\bf 1}_0$ & $ {\bf 4}_{-5} $ & \\
\hline \rule[-1mm]{0mm}{6mm} 6-form & ${\bf \overline{40}}$ &  &${\bf {4}}_7$ & $ {\bf 10}_{2} + {\bf 6}_2 $& ${\bf \overline{20}}_{-3} $ & \\
\cline{2-7} \rule[-1mm]{0mm}{6mm}  & ${\bf \overline{15}}$ &  & & ${\bf \overline{10}}_2$  & $ {\bf \overline{4}}_{-3} $&$ {\bf 1}_{-8}$\\
\hline \rule[-1mm]{0mm}{6mm} 7-form & ${\bf {70}}$ & & ${\bf
\overline{4}}_9$ & ${\bf 15}_{4}+ {\bf 1}_{4}$ & ${\bf {36}}_{-1}  +
{\bf 4}_{-1}$ &
${\bf 10}_{-6} $\\
\cline{2-7} \rule[-1mm]{0mm}{6mm}  & ${\bf {45}}$ & &  & ${\bf {15}}_{4}$ & ${\bf 20}_{-1} + {\bf 4}_{-1}$ & ${\bf 6}_{-6}$ \\
\cline{2-7} \rule[-1mm]{0mm}{6mm}  & ${\bf {5}}$ &  & & ${\bf 1}_4$ & ${\bf 4}_{-1}  $ & \\
 \hline
\end{tabular}
\end{center}
  \caption{\sl The decomposition of the $n$-form potentials of $D=7$ maximal
supergravity. The U-duality group is $\text{SL}(5, \mathbb{R})$ and
the T-duality group is $\text{SL}(4, \mathbb{R})$. We denote as a
subscript the $\mathbb{R}^+$-- charge (notation from
\cite{Slansky}). The value of $\alpha$ is given by $ \alpha =
\frac{1}{5} ( w -2n )$. \label{qD=7}}
\end{table}

 \begin{table}[t]
\begin{center}
\begin{tabular}{|c|c||c|c|c|c|c|c|}
\hline \rule[-1mm]{0mm}{6mm} field & U repr &  $\a=0$ & $\a=-1$ &
$\a=-2$ & $\a=-3$ & $\a=-4$ & $\a=-5$\\
\hline \hline \rule[-1mm]{0mm}{6mm} 1-form & ${\bf {16}}$ & $({\bf
8_{\rm C}})_{1}$ &
$({\bf 8_{\rm S}})_{-1}$ & &&&  \\
\hline \rule[-1mm]{0mm}{6mm} 2-form & ${\bf {10}}$ & ${\bf 1}_{2}$
 & $({\bf 8_{\rm V}})_{0}$ &
${\bf 1}_{-2}$&&&\\
\hline \rule[-1mm]{0mm}{6mm} 3-form & ${\bf \overline{16}}$ & & $({\bf 8_{\rm S}})_{1}$ & $({\bf 8_{\rm C}})_{-1}$&&&\\
\hline \rule[-1mm]{0mm}{6mm} 4-form & ${\bf {45}}$ & & $({\bf 8_{\rm V}})_{2}$ & ${\bf 28}_0 + {\bf 1}_0$ & $({\bf 8_{\rm V}})_{-2} $&&\\
\hline \rule[-1mm]{0mm}{6mm} 5-form & ${\bf {144}}$ & & $({\bf
8_{\rm S}})_{3}$ & $({\bf 8_{\rm C}})_{1}+({\bf 56_{\rm C}})_{1}$ &
$ ({\bf 8_{\rm S}})_{-1} + ({\bf 56_{\rm S}})_{- 1 }
$ & $({\bf 8_{\rm C}})_{-3}$ &\\
\hline \rule[-1mm]{0mm}{6mm} 6-form & ${\bf {320}}$ & &  $({\bf
{8}_{\rm V}})_{4}$ & $({\bf 35_{\rm V}})_{2}+ {\bf {28}}_{2}$ & $
2\times ({\bf 8_{\rm V}})_{0}  $ & $({\bf 35_{\rm
V}})_{- 2 }+ $ & $({\bf 8_{\rm V}})_{-4}$ \\
\rule[-1mm]{0mm}{6mm} && & & $+  {\bf {1}}_{2}$ & $+ ({\bf 160_{\rm
V}})_{0}$ &  ${\bf 28}_{-2} + {\bf 1}_{-2}
$&\\
\cline{2-8} \rule[-1mm]{0mm}{6mm}  & ${\bf \overline{126}}$ &  & & $
({\bf 35_{\rm S}})_{2} $ & $({\bf 56_{\rm
V}})_0$ & $({\bf 35_{\rm C}})_{-2 }$&  \\
\cline{2-8} \rule[-1mm]{0mm}{6mm}  & ${\bf 10}$ &  & & $  {\bf 1}_{2
}  $ & $({\bf
8_{\rm V}})_{0}$ & ${\bf 1}_{-2}$& \\
 \hline
\end{tabular}
\end{center}
  \caption{\sl The decomposition of the $n$-form potentials of $D=6$ maximal
supergravity. The U-duality is $\text{SO}(5,5)$, while the T-duality
is $\text{SO}(4,4)$. The relation between $\alpha$ and $w$ is  $
\alpha = \frac{1}{2} ( w -n)$. \label{qD=6}}
\end{table}

There are some general conclusions that one can draw by
looking at the representations occurring in Table \ref{NSformsanyD}.
First of all, the 6-forms have the peculiarity of always
containing the reducible representation of $\text{SO}(d, d)$ with
$d$ antisymmetric indices, which splits in self-dual and anti
self-dual irreducible representations. As we will see, this
corresponds to the fact that the world-volume multiplet for a
5-brane can either be a vector or a tensor multiplet. The
representations of the forms of rank higher than 6, which can only
occur in 7, 8 and 9 dimensions, always occur at least in pairs. In
nine dimensions the T-duality group is $\text{SO}(1,1)$, which means
that out of the fields in Table \ref{NSformsanyD}, only those with
at most 2 indices survive. The representations of the 7-forms are
two singlets, the representation of the 8-form is a vector which
decomposes in self-dual and anti self-dual representations, and
finally the 9-forms are three singlets. In eight dimensions the
T-duality is $\text{SO}(2,2)$, which means that the two
representations of the 7-forms are the same, while the
representation with four indices of the 8-form is a scalar and the
representation with two indices splits in self-dual and anti
self-dual. In seven dimensions the T-duality is $\text{SO}(3,3)$,
and thus the representation of the 7-form with two indices and the
one with four indices coincide.
%Above six dimensions, the lowest
%value of $\alpha$ is $-4$, while in six dimensions one can see from
%Table \ref{qD=6} that a 6-form with $\alpha=-5$ occurs. Similarly,
%one finds a 5-form with $\alpha=-5$ in five dimensions, as can be
%seen from Table \ref{qD=5}.

\begin{table}[t]
\begin{center}
\begin{tabular}{|c|c||c|c|c|c|c|c|}
\hline \rule[-1mm]{0mm}{6mm} field & U repr &  $\a=0$ & $\a=-1$ &
$\a=-2$ & $\a=-3$ & $\a=-4$ & $\a=-5$\\
\hline \hline \rule[-1mm]{0mm}{6mm} 1-form & ${\bf {27}}$ & ${\bf
{10}}_{-2}$ &
${\bf {16}}_{1}$ & ${\bf 1}_{4}$ & & & \\
\hline \rule[-1mm]{0mm}{6mm} 2-form & ${\bf \overline{27}}$ & ${\bf
1}_{-4}$ & ${\bf \overline{16}}_{-1}$ &
 ${\bf 10}_2$ & & &\\
\hline \rule[-1mm]{0mm}{6mm} 3-form & ${\bf {78}}$  & & ${\bf {16}}_{-3}$ &  ${\bf 45}_0 + {\bf 1}_0$ & ${\bf \overline{16}}_{3}$ & & \\
\hline \rule[-1mm]{0mm}{6mm} 4-form & ${\bf 351}$ & & ${\bf
\overline{16}}_{-5}$ & ${\bf 120}_{-2}+  { \bf 10}_{-2}$ & ${\bf
16}_{1} + {\bf 144}_1 $ & ${\bf 45}_4$ & \\
\hline \rule[-1mm]{0mm}{6mm} 5-form & ${\bf \overline{1728}}$ & &
${\bf {16}}_{-7}$ & ${\bf 1}_{-4} + {\bf 45}_{-4}$ &  $2 \times{\bf
\overline{16}}_{-1}+$ & ${\bf 10}_2 + {\bf 120}_2 +$ & ${\bf 144}_5$\\
\rule[-1mm]{0mm}{6mm} && & &  $+ {\bf {210}}_{-4}$ & ${\bf
\overline{144}}_{-1} + {\bf \overline{560}}_{-1}$ & ${\bf 320}_2 +
{\bf \overline{126}}_2 $ &
\\
\cline{2-8} \rule[-1mm]{0mm}{6mm}  & ${\bf \overline{27}}$ &  & & $
{\bf 1}_{-4}$ &
${\bf \overline{16}}_{-1}$ & $ {\bf 10}_2 $ & \\
 \hline
\end{tabular}
\end{center}
  \caption{\sl The decomposition of the $n$-form potentials of $D=5$ maximal
supergravity. The U-duality is $\text{E}_6$ and the T-duality is
SO(5,5). The relation between $\alpha$ and the weight $w$ is given
by the equation $\alpha = -\frac{1}{3} ( w +2 n)$. \label{qD=5}}
\end{table}

\section{String solitons in ten dimensions}

The purpose of this section is to consider the solitonic $D$-fields
of IIA and IIB supergravity and to investigate which of these fields
can couple to a solitonic brane with a worldvolume supersymmetric
field content. We have summarised these $D$-fields in Table
\ref{IIAIIBD}. The results of this investigation will not be
surprising: we will find that the only solitonic branes are the
NS5A-brane, with a worldvolume tensor multiplet, and the
 NS5B-brane, with a worldvolume vector multiplet. More precisely,
only one of the 4 solitonic  IIA (IIB) $D$-fields in Table
\ref{IIAIIBD} corresponds to a supersymmetric solitonic IIA (IIB)
brane.

An important lesson we learn from this analysis is that, even in ten
dimensions, the requirement of a worldvolume supersymmetric field
content excludes most of the $D$-fields as potentials suggesting new
solitonic branes. The status of these $D$-fields at present is
unclear. As already stressed in the introduction this does {\sl not}
necessarily imply that these $D$-fields have nothing to do with
branes at all.  In this paper, however, we will restrict our
attention to the solitonic branes of string theory that have a known
supersymmetric worldvolume field content.

The restriction we find on the allowed solitonic $D$-fields results
from a clash between the known supersymmetric worldvolume multiplets
with 16 supercharges, i.e.~the vector multiplet and the tensor
multiplet, and the number of worldvolume fields that we need to
construct a gauge-invariant Wess-Zumino term. Since the tensor
multiplet only exists in 6 dimensions, only 5-branes can have a
worldvolume dynamics governed by a tensor multiplet. Both
fundamental objects as well as D-branes can end on the Solitonic
Brane and this leads to many worldvolume $p$-form potentials. It is
non-trivial to fit all these potentials into a vector or tensor
multiplet. Since vector multiplets already occur in the case of
D-branes we expect these solitonic branes to be the S-duals of
D-branes. In contrast, we expect the solitonic 5-branes with tensor
multiplets to be related, via dimensional reduction, to the M5-brane
of M-theory which also has a worldvolume tensor multiplet.

We now consider the occurrence of worldvolume $p$-form fields for
the ten-dimensional solitonic branes in more detail and show in
which cases they fit a vector or tensor  multiplet. Both IIA and IIB
supergravity contain the Fundamental 2-form potential $B_{2}$, with
curvature $H_{3}$, that couples to the Fundamental String. They
differ however in the RR-potentials in the sense that IIA (IIB)
supergravity contain odd-form (even-form) RR potentials. All RR
potentials couple to the D-branes. Using the same compact notation
as in \cite{Bergshoeff:2010xc} we can write the transformation rules
of the IIA and IIB RR potentials and the expressions for their
curvatures as
  \begin{equation}
  \delta C = d \lambda + H_3 \lambda
  \end{equation}
and
  \begin{equation}
  G = d C + H_3 C\,,
  \end{equation}
respectively. Here $\lambda$ denote the RR gauge parameters. It is
understood that in the IIA case we pick out the odd-form potentials
whereas in the IIB case we take the even-form potentials. One then
introduces a worldvolume 1-form field $b_1$ associated to the
fundamental 2-form $B_2$, whose gauge-invariant curvature is
  \begin{equation}
  {\cal F}_2 = d b_1 +B_2 \quad ,
  \end{equation}
and constructs a gauge-invariant WZ-term which was written in the
introduction in eq. \eqref{WZterm}.

Using the same short-hand notation the gauge transformations of the
solitonic $D$-fields can be written as (here $G\lambda$ denotes the
various contributions of the right rank; we will give precise
coefficient of this term in the two subsections below)
\begin{equation}
\delta D = d\Lambda + G\lambda + H_3\Lambda\,,
\label{10dimgaugetransfDfieldgeneral}
\end{equation}
where $\Lambda$ denote the solitonic gauge parameters. The
corresponding gauge-invariant curvatures are given by
\begin{equation}
F = dD +  GC + H_3D\, , \label{10dimfieldstrengthDfieldgeneral}
\end{equation}
where again $GC$ schematically denotes the various terms such that
the rank of $G$ plus the rank of $C$ equals the rank of $F$.

Given that the RR field strengths occur in the gauge transformations
of the $D$ fields, in order to write down a gauge-invariant WZ term
one introduces worldvolume fields $c$ associated to the RR fields.
These worldvolume fields are even forms for IIA and odd-forms for
IIB. Note that this is the opposite of the target-space potentials.
The gauge-invariant worldvolume curvatures for the $c$ potentials
are given by
\begin{equation}
{\cal G}= dc + H_3 c + C\,,
\end{equation}
with $\delta c = -\lambda$. They satisfy the Bianchi identities
\begin{equation}
d{\cal G} = -H_3\,{\cal G}+G\,.
\end{equation}

Using the above formulae one can show that the following candidate
WZ-term is (trivially) gauge-invariant:
\begin{equation}
{\cal L}_{\text{WZ}}(\text{solitonic}) = e^{{\cal F}_2}\left ( D +
{\cal G}C\right )\,.
\end{equation}
The danger of this Ansatz is that it contains many worldvolume
potentials hidden in the worldvolume curvatures ${\cal F}_2$ and
${\cal G}$ and they are difficult to fit into a vector or tensor
multiplet.

Below we will scan the different cases for IIA and IIB supergravity
separately, give the precise formulae, and see in which cases we can
put all worldvolume fields into a given multiplet.

\subsection{The IIB solitonic WZ terms}

We first consider the 6-form potential $D_6$ corresponding to the
NS5B-brane. The gauge transformations and curvatures are given by
  \begin{equation}
  \delta D_6 = d \Lambda_5 +  G_5 \lambda_1 - G_3 \lambda_3
  \end{equation}
and
  \begin{equation}
  F_7 = d D_6 +  G_5 C_2 -  G_3 C_4\,,
  \end{equation}
respectively. The gauge-invariant  WZ term is given by
\begin{equation}
{\cal L}_{\text{WZ}} = D_6 +  {\cal G}_4C_2-  {\cal G}_2C_4 \equiv
{\cal D}_6\,,
\end{equation}
which contains the worldvolume potentials $c_1$ and $c_3$. Imposing
a duality relation between them we are left with one vector.
Together with the 4 embedding scalars these form the 8 bosonic
degrees of freedom of an 8+8 vector multiplet in the six-dimensional
worldvolume.

\begin{table}[t]
\begin{center}
\begin{tabular}{|c|c|c|}
\hline
 field& IIA brane/multiplet&IIB brane/multiplet\\[.1truecm]
\hline \hline \rule[-1mm]{0mm}{6mm} $D_6$&NS5A/tensor multiplet & NS5B/vector multiplet\\[.1truecm]
\hline \rule[-1mm]{0mm}{6mm} $D_8$&--& -- \\[.1truecm]
\hline \rule[-1mm]{0mm}{6mm} $D_{10}\,, D_{10}^\prime$&--& --\\[.1truecm]
 \hline
\end{tabular}
\end{center}
  \caption{\sl The solitonic $D$-fields of IIA and IIB supergravity. The second and third column
  indicate which $D$-fields couple to the known  solitonic branes of string theory.
  The worldvolume multiplets of these branes are indicated. }
  \label{IIAIIBD}
\end{table}

We next consider the solitonic  8-form $D_8$. An important
difference with the previous case is that $D_8$ transforms to $H_3$
under a solitonic gauge transformation. Basically, this means that
fundamental strings can end on this soliton and, hence, there exists
a corresponding Born-Infeld vector $b_1$ in the worldvolume. As we
will see, this will lead to too many worldvolume fields to fit a
multiplet. Partially, this effect is cancelled (but not enough) in
this case due to the fact that the transformation rule and curvature
contain a free real parameter $\alpha$. Indeed, the general form of
the gauge transformation and field strength in eqs.
\eqref{10dimgaugetransfDfieldgeneral} and
\eqref{10dimfieldstrengthDfieldgeneral}, requiring gauge invariance
of the field strength leads to
  \begin{equation}
  \delta D_8 = d \Lambda_7 + (1-\alpha )  G_7 \lambda_1 +\alpha  G_5 \lambda_3 -
  (1+ \alpha)  G_3 \lambda_5 + H_3 \Lambda_5
  \label{8formIIBwithalpha}
  \end{equation}
and
  \begin{equation}
  F_9 = dD_8 + (1- \alpha )  G_7 C_2 + \alpha  G_5 C_4 - (1+ \alpha )  G_3 C_6 + H_3 D_6\,.
  \end{equation}
The free parameter $\alpha$ can be reabsorbed by the redefinition
\begin{equation}
D_8\ \rightarrow\ D_8  + \alpha (C_2 C_6 - \tfrac{1}{2} C_4 C_4 )\,.
\end{equation}
That this is an allowed redefinition can be seen from the fact that
 \begin{equation}
 d (C_2 C_6 - \tfrac{1}{2} C_4 C_4 ) = G_3 C_6 + G_7 C_2 - G_5 C_4\,
 ,
 \end{equation}
which means that the terms $H_3 CC$ vanish, and therefore adding
this term preserves the structure of eq.
\eqref{10dimfieldstrengthDfieldgeneral}. A gauge-invariant WZ term
for this case is given by
\begin{equation}
{\cal L}_{\text{WZ}} = D_8 + (1- \alpha){\cal G}_6C_2 + \alpha {\cal
G}_4C_4 - (1+ \alpha){\cal G}_2C_6 + {\cal F}_2{\cal D}_6\equiv
{\cal D}_8 + {\cal F}_2{\cal D}_6\,,
\end{equation}
which contains the worldvolume fields $c_1,c_3,c_5$ and $b_1$. The
problem is now that there are too many worldvolume fields to fit
into a multiplet. Even if we impose a duality relation between $c_1$
and $c_5$ we are left with two vectors and one 3-form potential. The
free parameter $\alpha$ may be used to eliminate one of the vectors
or the 3-form potential but not both. Hence, for no choice of
$\alpha$ will the worldvolume fields fit into a vector multiplet. We
can without loss of generality choose a value for the parameter
$\alpha$, and we choose it to vanish, so that the field strength and
gauge transformation are symmetric with respect to electromagnetic
duality on the worldvolume, that is
  \begin{equation}
  \delta D_8= d \Lambda_7 +   G_7 \lambda_1 -
  G_3 \lambda_5 + H_3 \Lambda_5 \quad . \label{8formIIBwithoutalpha}
  \end{equation}

Finally, we consider the 10-form potentials. These are determined by
formally requiring that their field strength is gauge invariant,
although this is actually identically zero. Taking $\alpha=0$ in eq.
\eqref{8formIIBwithalpha}, that is using eq.
\eqref{8formIIBwithoutalpha}, one obtains
  \begin{equation}
  \delta D_{10} = d \Lambda_9 + \beta G_9 \lambda_1 + (1 -\beta) G_7
  \lambda_3 - ( 1 -\beta ) G_5 \lambda_5 - \beta G_3 \lambda_7 +
  H_3 \Lambda_7 \quad ,
  \end{equation}
where $\beta$ is an arbitrary real parameter. Given that no field
redefinition is possible in this case, this parameter is the
relative normalisation of the two independent 10-forms. The
corresponding gauge-invariant WZ term is given by
\begin{equation}
{\cal L}_{\text{WZ}} = D_{10} + \beta {\cal G}_8 C_2 + (1 - \beta )
{\cal G}_6 C_4 - ( 1 -\beta) {\cal G}_4 C_6 - \beta {\cal G}_2 C_8
 + {\cal F}_2{\cal D}_8 + \tfrac{1}{2}{\cal F}_2{\cal F}_2{\cal
 D}_6\, .
\end{equation}
This WZ term contains the potentials $c_1,c_3,c_5,c_7$ and $b_1$,
which after imposing duality relations on the $c$ fields corresponds
to a vector (multiplied by $\beta$) and a 3-form (multiplied by
$1-\beta$) plus another scalar. For no choice of the parameter
$\beta$ can the field content be the bosonic sector of a
ten-dimensional vector multiplet.

\subsection{The IIA solitonic WZ terms}

We next consider the solitonic potentials of IIA supergravity (for
zero Romans mass parameter). We first consider the 6-form potential
$D_6$ corresponding to the NS5A brane. The gauge transformations and
curvature for this case are given by
  \begin{equation}
  \delta D_6 = d \Lambda_5 + G_6 \lambda_0 - G_4 \lambda_2 + G_2
  \lambda_4
  \end{equation}
and
  \begin{equation}
  F_7 = d D_6 - G_6 C_1 + G_4 C_3 - G_2 C_5\,,
   \end{equation}
respectively. The corresponding gauge-invariant WZ term is given by
\begin{equation}
{\cal L}_{\text{WZ}} = D_6 - {\cal G}_5 C_1 + {\cal G}_3C_3 -{\cal
G}_1C_5 \equiv {\cal D}_6\,,
\end{equation}
which contains the worldvolume potentials $c_0, c_2$ and $c_4$.
Imposing a duality relation between them we are left with a scalar
and a self-dual 2-form. Together with the 4 embedding scalars these
form the bosonic sector of an 8+8 tensor multiplet in the
six-dimensional worldvolume. The extra $c_0$ scalar indicates that
this solitonic 5-brane has an 11-dimensional origin as the M5-brane.

We next consider the 8-form $D_8$. Like in the IIB case $D_8$
transforms to $H_3$ under a solitonic gauge transformation.
Furthermore, the transformation rules and curvature contain a free
real parameter $\alpha$:
  \begin{equation}
  \delta D_8 = d \Lambda_7 + \alpha G_8 \lambda_0 + (1-\alpha) G_6 \lambda_2
  -(2 - \alpha )
  G_4 \lambda_4 + (3-\alpha ) G_2 \lambda_6 + H_3 \Lambda_5
  \end{equation}
and
  \begin{equation}
  F_9 = dD_8 -\alpha G_8 C_1 - (1 -\alpha ) G_6 C_3 + (2 -\alpha )
  G_4 C_5 -  (3 -\alpha ) G_2 C_7 + H_3 D_6 \quad .
  \end{equation}
As in the IIB case the presence of the free parameter $\alpha$ is
related to the field redefinition
  \begin{equation}
  D_8\ \rightarrow\ D_8 + \alpha (C_1 C_7 -  C_3 C_5 )\,.
  \end{equation}
This a consistent redefinition due to the fact that
 \begin{equation}
 d (C_1 C_7 -  C_3 C_5 ) = G_2 C_7 - G_4 C_5 + G_6 C_3 - G_8 C_1\,.
 \end{equation}
The gauge-invariant WZ term is given by
\begin{equation}
{\cal L}_{\text{WZ}} = D_8 -\alpha {\cal G}_7C_1-(1 -\alpha){\cal
G}_5C_3 +(2-\alpha ) {\cal G}_3C_5 -(3-\alpha) {\cal G}_1C_7 + {\cal
F}_2{\cal D}_6\equiv {\cal D}_8 + {\cal F}_2{\cal D}_6 \,,
\end{equation}
which contains the worldvolume fields $c_0,c_2,c_4,c_6$ and $b_1$.
Imposing a duality relation on the $c$ fields we are left with a
scalar, a (non-self-dual) 2-form and a vector. For no choice of
$\alpha$ can we get rid of both the scalar and the 2-form and
therefore these fields will not fit into a vector multiplet. As in
the IIB case, there is a choice of $\alpha$ for which the gauge
transformations look the most symmetric. This choice is $\alpha =
3/2$, which gives
  \begin{equation}
  \delta D_8 = d \Lambda_7 + \tfrac{3}{2} G_8 \lambda_0 -\tfrac{1}{2} G_6 \lambda_2
  - \tfrac{1}{2}
  G_4 \lambda_4 + \tfrac{3}{2} G_2 \lambda_6 + H_3 \Lambda_5 \quad .
  \label{8formIIAsymmetric}
  \end{equation}

Finally, we consider the  two 10-form potentials. In this case one
obtains that the transformations depend on a real parameter $\beta$,
and in particular using eq. \eqref{8formIIAsymmetric} one gets
  \begin{eqnarray}
  & & \delta D_{10} = d \Lambda_9 + \beta G_{10} \lambda_0 +
  (\tfrac{3}{2} - \beta ) G_8 \lambda_2 - (2 -\beta ) G_6 \lambda_4
  +   (\tfrac{3}{2} - \beta ) G_4 \lambda_6 \nonumber \\
  & & \quad \qquad + \beta G_2 \lambda_8 +
  H_3 \Lambda_7\, .
  \end{eqnarray}
Given that no field redefinition is possible, the parameter $\beta$
signals the presence of two independent 10-forms. The corresponding
gauge-invariant WZ term is given by
\begin{eqnarray}
{\cal L}_{\text{WZ}} &=& D_{10}  - \beta {\cal G}_9 C_1
 - (\tfrac{3}{2} -\beta ) {\cal G}_7 C_3 + (2 -\beta ) {\cal G}_5
 C_5 - (\tfrac{3}{2} -\beta ) {\cal G}_3 C_7 - \beta {\cal G}_1 C_9
 \nonumber \\
 & &
 + {\cal F}_2{\cal D}_8 + \tfrac{1}{2}{\cal F}_2{\cal F}_2{\cal
 D}_6\, ,
\end{eqnarray}
which contains the potentials $c_0,c_2,c_4,c_6,c_8$ and $b_1$. This
corresponds to a scalar multiplied by $\beta$, a 2-form multiplied
by $\tfrac{3}{2} -\beta $, a self-dual 4-form multiplied by $2
-\beta$ plus a vector, and for no choice of the parameter $\beta$
one obtains the bosonic sector of a ten-dimensional vector
multiplet.

\section{Gauge algebra of solitonic fields in any dimension}
In this section we want to determine the gauge algebra of the fields
in Table \ref{NSformsanyD}. In \cite{Bergshoeff:2010xc} the gauge
algebra of the $\alpha =0$ and $\alpha =-1$ fields in any dimension
was considered. Because of conservation of the $\mathbb{R}^+$
weight, the $\alpha =-2$ fields, which are the $D$ fields, transform
among themselves and into the $B$ fields and $C$ fields of
\cite{Bergshoeff:2010xc}, but not into fields with lower $\alpha$.
In order to proceed, we first review the results and the notation of
\cite{Bergshoeff:2010xc}. The $\alpha=0$ fields are $B_{1,A}$ and
$B_2$, and their corresponding gauge transformations and fields
strengths are
  \begin{eqnarray}
 & &  \delta B_{1,A} = d \Sigma_{0,A} \nonumber \\
 & & \delta B_2 = d \Sigma_1 - H_{2,A} \Sigma_0^A
 \label{gaugealgebraalpha=0rev}
 \end{eqnarray}
 and
  \begin{eqnarray}
  & & H_{2,A} = d B_{1,A} \nonumber \\
  & & H_3 = d B_2 + H_{2,A} B_1^A \quad .
  \end{eqnarray}
This implies the Bianchi identity
 \begin{equation}
 d H_3 = H_{2,A} H_2^A \quad .
 \end{equation}
The $\alpha=-1$ fields $C$ are spinors of $\text{SO}(d,d)$, of plus
or minus chirality according to whether the form has odd or even
rank. That is, following the notation explained in appendix A,
  \begin{equation}
  C_{2n+1, \alpha} =
\begin{pmatrix}
C_{2n+1, a}\\
0
\end{pmatrix} \quad , \qquad C_{2n, \alpha} =
\begin{pmatrix}
0\\
C_{2n, \dot{a}}
\end{pmatrix} \quad .
\end{equation}
Similarly, the gauge parameters are
  \begin{equation}
  \lambda_{2n, \alpha} =
\begin{pmatrix}
\lambda_{2n, a}\\
0
\end{pmatrix} \quad , \qquad \lambda_{2n-1, \alpha} =
\begin{pmatrix}
0\\
\lambda_{2n-1, \dot{a}}
\end{pmatrix} \quad .
\end{equation}
The gauge transformations and field strengths are
  \begin{equation}
  \delta C_{n, \alpha} = d \lambda_{n-1, \alpha} + H_3
  \lambda_{n-3,\alpha} - H_{2,A} \Gamma^A_\alpha{}^\beta
  \lambda_{n-2,\beta} \label{gaugealgebraalpha=-1rev}
  \end{equation}
and
  \begin{equation}
  G_{n+1,\alpha} = d C_{n,\alpha} + H_3 C_{n-2,\alpha} + H_{2,A} \Gamma^A_\alpha{}^\beta
  C_{n-1,\beta} \quad .
  \end{equation}
This implies
  \begin{equation}
   d G_{n+1,\alpha} = -H_3 G_{n-1,\alpha} + H_{2,A} \Gamma^A_\alpha{}^\beta
  G_{n,\beta} \quad .
  \end{equation}
In the following we will consider bilinears made out of these
$\text{SO}(d,d)$ spinors. We thus define $\overline{G}$ as
  \begin{equation}
  \overline{G}^\alpha = i G_\beta C^{\beta \alpha} \ \ {\rm for} \ \  d =2,3 \qquad
    \overline{G}^\alpha = G_\beta C^{\beta \alpha} \ \ {\rm for} \ \ d
    =1,4,5 \quad ,
    \end{equation}
where $C^{\alpha \beta}$ is the charge conjugation matrix, whose
properties for all $d$ are described in appendix A. In the following
we will also need the relations
  \begin{equation}
  \overline{G}_{n+1}^\alpha = d  \overline{C}_n^\alpha +  H_3 \overline{C}_{n-2}^\alpha +
  H_{2,A} \overline{C}_{n-1}^\beta \Gamma^A_\beta{}^\alpha
  \end{equation}
and
  \begin{equation}
  d \overline{G}_{n+1}^\alpha = - H_3 \overline{G}_{n-1}^\alpha +
  H_{2,A} \overline{G}_{n}^\beta \Gamma^A_\beta{}^\alpha \quad
  ,\label{dGbarisHGbar}
  \end{equation}
where we have made use of eq. \eqref{lambdabarCrelation}. The
bilinears in our expressions will always have all their spinor
indices contracted, and we will always write them without indices,
so that the contraction will always be understood.

We now proceed with deriving the gauge algebra of the $D$ fields,
which are listed in Table \ref{NSformsanyD}. We denote the gauge
parameters of the $D$ fields with $\Lambda$, and the field strengths
with $F$. We want to determine the gauge algebra of the $D$ fields
in the abelian basis, that is we require that the gauge
transformations only contain gauge invariant quantities. Up to field
redefinitions, the gauge transformations can then always be written
schematically as $ \delta D = d \Lambda + \overline{G} \Gamma
\lambda + H \Lambda $, where here $\Gamma$ schematically denotes the
antisymmetric product of a number of gamma matrices equal to the
number of $\text{SO}(d,d)$ indices of the field $D$. Before one
determines the gauge transformations of the $D$ fields, one has to
be sure that the abelian basis that one chooses is not redundant,
that is that there are no field redefinitions within the basis. What
we are looking after is the possibility that a bilinear
$\overline{C} \Gamma C$ in the $C$ fields is such that
$d(\overline{C} \Gamma C) = G \Gamma C$. When this occurs, we have
to subtract this to the number of $D$ fields that close the algebra
in the abelian basis to get the actual counting right. We want to
determine an $m$-form given by the combination
  \begin{equation}
  \sum_{n=1}^{\left[\frac{m}{2}\right]} \alpha_n \overline{C}_{m-n}
  \Gamma_{A_1 ...A_p}
  C_n
  \end{equation}
such that when computing its curl all terms of the form $H
\overline{C} \Gamma C$ vanish. Given the chirality of the fields and
the form of the $C$ matrix in any $d$, given in eq.
\eqref{Cmatrixforminanyd}, the only possibility is that $m+p +D$
must be even. We first consider the case $p=0$, that is the case of
a T-duality singlet. By direct inspection, one can show that in any
$D$ there is a solution for $m=D-2$, and the whole set of solutions
is
  \begin{eqnarray}
  & & m=2: \qquad \overline{C}_1 C_1 \qquad \qquad \qquad \quad \qquad \  (D=4,8)\nonumber \\
  & & m=3: \qquad \overline{C}_2 C_1 \qquad \qquad \qquad \quad \qquad \  (D=5,9)\nonumber \\
  & & m=4: \qquad \overline{C}_3 C_1 - \tfrac{1}{2} \overline{C}_2
  C_2 \qquad \qquad \quad  \  (D=6) \nonumber \\
  & & m=5 : \qquad \overline{C}_4 C_1 + \overline{C}_3 C_2 \qquad
  \qquad \quad  \quad
  (D=7) \nonumber \\
  & & m=6 : \qquad \overline{C}_5 C_1 - \overline{C}_4 C_2
  -\tfrac{1}{2} \overline{C}_3 C_3 \qquad (D=8) \nonumber \\
  & & m=7 : \qquad \overline{C}_6 C_1 + \overline{C}_5 C_2 -
  \overline{C}_4 C_3 \qquad \ \  (D=9)\label{redundancyCC}
  \end{eqnarray}
One can then show that for $p \neq 0$ there are no solutions. To
summarise, this analysis shows that this field redefinition will
only affect the counting of the $D-2$-forms that are singlets of
$\text{SO}(d,d)$. By looking at Table \ref{NSformsanyD}, we expect
to find a single field $D_{D-2}$ after this field redefinition is
taken into account. As we will see, this will be confirmed by our
explicit computations.

We now determine the gauge algebra of the fields given in Table
\ref{NSformsanyD}. We start by analysing the fields with the highest
amount of T-duality indices for each rank. These fields are
  \begin{equation}
  D_{D-4+m, A_1 ...A_m} \qquad m=0,1,...,4
  \label{DD-4+mfieldsforanym}
  \quad ,
  \end{equation}
and one can see that the redundancy of eq. \eqref{redundancyCC} does
not affect these fields. We write the gauge transformation of these
fields as
  \begin{eqnarray}
  & & \delta D_{D-4+m, A_1 ...A_m} = d \Lambda_{D-5+m, A_1 ...A_m}  +
  \sum_{n=0}^{D-6+m} a_n^{(m)} \overline{G}_{D-4+m-n} \Gamma_{A_1
  ...A_m} \lambda_n \nonumber \\
  & & \qquad - m H_{2,[A_1} \Lambda_{D-6+m, A_2 ...A_m ]}
  \label{variationDD-4+m}
  \end{eqnarray}
and the corresponding field strength as
  \begin{eqnarray}
  & & F_{D-3+m, A_1 ...A_m} = d D_{D-4+m, A_1 ...A_m}  +
  \sum_{n=0}^{D-6+m} b_n^{(m)} \overline{G}_{D-4+m-n} \Gamma_{A_1
  ...A_m} C_{n+1} \nonumber \\
  & & \qquad + m H_{2,[A_1} D_{D-5+m, A_2 ...A_m ]} \quad .
  \label{fieldstrengthDD-3+m}
  \end{eqnarray}
The gauge invariance with respect to the gauge parameters $\Lambda$
and $\Sigma$ is already implied by the form in which we have written
the last term in eqs.~\eqref{variationDD-4+m} and
\eqref{fieldstrengthDD-3+m}, while the real parameters $a_n^{(m)}$
and $b_n^{(m)}$ can be determined by imposing gauge invariance of
the curvatures $F_{D-3+m,A_1 ...A_m}$ with respect to the parameters
$\lambda$.  Formally, this applies to the $D$-forms as well,
although their field strengths are actually identically zero. Using
equation \eqref{dGbarisHGbar}, the reader can see that the variation
of $F_{D-3+m, A_1 ...A_m}$ produces terms of four different
structures, that is the $d \lambda$ terms $\overline{G} \Gamma_{A_1
...A_m} d \lambda$, two types of $H_2$ terms $H_2^B \overline{G}
\Gamma_{BA_1 ...A_m} \lambda$ and $H_{2, [A_1} \overline{G}
\Gamma_{A_2 ...A_m ]} \lambda$, and finally the $H_3$ terms $H_3
\overline{G} \Gamma_{A_1 ...A_m} \lambda$. Putting to zero all of
them one obtains the equations
  \begin{eqnarray}
  & & a_n^{(m)} (-)^{D+m-n} + b_n^{(m)} = 0 \quad \qquad \quad
  n=0,...,D-6+m \nonumber \\
   && a_n^{(m)} - b_{n+1}^{(m)} (-)^m = 0 \quad \qquad \quad \ \qquad n=0,...,D-7+m
   \nonumber \\
   & & m ( a_n^{(m)} + b_{n+1}^{(m)} (-)^m + a_n^{(m-1)}) =0 \quad
   n=0,...,D-7+m \nonumber \\
  & & -a_n^{(m)} + b_{n+2}^{(m)} (-)^{D+m-n} = 0 \quad \quad \quad \
  n=0,...,D-8+m \quad .
   \end{eqnarray}
Fixing the normalisation so that $a_0^{(0)} =1$, the solution of
this set of equations is
  \begin{eqnarray}
  & & a_0^{(m)} = \left( -\tfrac{1}{2} \right)^m \nonumber \\
  & & a_1^{(m)} = \left( -\tfrac{1}{2} \right)^m  (-)^D \nonumber \\
  & & b_0^{(m)} = - \left( \tfrac{1}{2} \right)^m (-)^D \nonumber \\
  & & b_1^{(m)} = \left( \tfrac{1}{2} \right)^m
  \label{valuesofavaluesofb}
  \end{eqnarray}
while all other values are given by the relation
  \begin{equation}
  a_{n+2}^{(m)} + a_n^{(m)} = 0 \qquad b_{n+2}^{(m)} + b_n^{(m)} =0
  \quad . \label{andbrecursion}
  \end{equation}
This gives the gauge algebra of all the fields in equation
\eqref{DD-4+mfieldsforanym}.

Next, we consider the other fields in Table \ref{NSformsanyD}, that
are $D_{D-2}$, $D_{D-1,A}$ and $D_D$, $D_D^\prime$ and $D_{D,AB}$.
We first consider the fields
  \begin{equation}
  D_{D-2+ m , A_1 ... A_m} \qquad \quad m=0,1,2 \quad ,
  \end{equation}
while the $D$-form singlets will be separately discussed later. We
write the gauge transformation as
  \begin{eqnarray}
  & & \delta D_{D-2+m ,A_1 ...A_m} = d \Lambda_{D-3+m, A_1 ...A_m} +
  \sum_{n=0}^{D-4+m} c_n^{(m)} \overline{G}_{D-2+m-n}
  \Gamma_{A_1...A_m} \lambda_n \nonumber \\
  & & + H_3 \Lambda_{D-5+m , A_1 ...A_m} -
  m H_{2 ,[A_1} \Lambda_{D-4+m, A_2 ...A_m]} - H_2^B \Lambda_{D-4+m,
  BA_1 ...A_m}
  \end{eqnarray}
while the field strength is
  \begin{eqnarray}
  & & F_{D-1+m ,A_1 ...A_m} = d D_{D-2+m, A_1 ...A_m} +
  \sum_{n=0}^{D-4+m} d_n^{(m)} \overline{G}_{D-2+m-n}
  \Gamma_{A_1...A_m} C_{n+1} \nonumber \\
  & & + H_3 D_{D-4+m , A_1 ...A_m} +
  m H_{2 ,[A_1} D_{D-3+m, A_2 ...A_m]} + H_2^B D_{D-3+m,
  BA_1 ...A_m} \quad ,
  \end{eqnarray}
where $c_n^{(m)}$ and $d_n^{(m)}$ are real parameters. Again, we
have written the field strengths and gauge transformations so that
gauge invariance of the field strengths with respect to the
parameters $\Sigma $ and $\Lambda$ is already implied, while
imposing gauge invariance with respect to $\lambda$ leads to the
equations
  \begin{eqnarray}
  & & c_n^{(m)} (-)^{D+m-n} + d_n^{(m)} = 0 \qquad \quad \qquad
  \quad
  n=0,...,D-4+m \nonumber \\
  && c_n^{(m)} - d_{n+1}^{(m)} (-)^m + a_n^{(m+1)} =0 \qquad \qquad
  n=0,...,D-5+m \nonumber \\
  & & m ( c_n^{(m)} + d_{n+1}^{(m)} (-)^m + c_n^{(m-1)} ) = 0 \qquad
  \
  n = 0 ,...,D-5+m \nonumber \\
  & & - c_n^{(m)} + d_{n+2}^{(m)} (-)^{D+m-n} + a_n^{(m)} =0 \qquad
  n = 0 ,..., D-6+m
  \quad . \label{solutionforcd}
  \end{eqnarray}
The reader can realise by direct inspection that there is a one
parameter family of solutions in each dimension. This is expected
from eq.~\eqref{redundancyCC}, which shows that in each dimension
one can add to the $(D-2)$-form a suitable combination quadratic in
the $C$ fields such that the gauge transformation of the resulting
field is still in the abelian basis. Taking into account this
redundancy, one  finds the solutions corresponding to the $D_{D-2+m,
A_1 ...A_m}$ as in Table \ref{NSformsanyD}.

We consider explicitly here as a prototype example the
five-dimensional case. Denoting with $\alpha$ the undetermined
parameter, we have that the algebra of the $D_3$ form closes if
  \begin{eqnarray}
  & &c_0^{(0)} = \alpha \quad \qquad c_1^{(0)} = -\alpha +
  \tfrac{1}{2} \nonumber \\
  & & d_0^{(0)} = \alpha \quad \qquad d_1^{(0)} = \alpha -
  \tfrac{1}{2} \quad . \label{3forminfive}
  \end{eqnarray}
We choose to fix the value of $\alpha$ such that the solution looks
the most symmetric. This value is $\alpha = \tfrac{1}{4}$, as can be
seen from eq. \eqref{3forminfive}. This gives
  \begin{eqnarray}
  & &c_0^{(0)} = \tfrac{1}{4} \quad \qquad c_1^{(0)} =
  \tfrac{1}{4} \nonumber \\
  & & d_0^{(0)} = \tfrac{1}{4} \quad \qquad d_1^{(0)} = -
  \tfrac{1}{4} \quad . \label{3forminfivefixedalpha}
  \end{eqnarray}
Plugging this solution back in equation \eqref{solutionforcd} one
then finds
  \begin{eqnarray}
  & & c_0^{(1)} = - \tfrac{1}{4} \qquad c_1^{(1)} = 0 \qquad
  c_2^{(1)} = -\tfrac{1}{4} \nonumber \\
  & & d_0^{(1)} =  \tfrac{1}{4} \qquad \ \ d_1^{(1)} = 0 \qquad
  d_2^{(1)} = \tfrac{1}{4} \label{4forminfivefixedalpha}
  \end{eqnarray}
for the $D-1$-form and
  \begin{eqnarray}
    & & c_0^{(2)} =  \tfrac{3}{16} \qquad c_1^{(2)} = - \tfrac{1}{16} \qquad
    c_2^{(2)} = \tfrac{1}{16} \qquad c_3^{(2)} = - \tfrac{3}{16} \nonumber \\
  & & d_0^{(2)} =  \tfrac{3}{16} \qquad d_1^{(2)} = \tfrac{1}{16}
  \qquad \ \
  d_2^{(2)} = \tfrac{1}{16} \qquad d_3^{(2)} = \tfrac{3}{16}
  \end{eqnarray}
for the $D$-form. Similarly, it is straightforward to derive the
solutions of eq. \eqref{solutionforcd} in other dimensions.

Finally, we consider the two $D$-form singlets in Table
\ref{NSformsanyD}. We write the gauge transformations and the fields
strength (again, the gauge transformations for these fields can be
determined imposing the formal gauge invariance of a field strength,
although this is actually identically zero) as
   \begin{equation}
   \delta D_D = d \Lambda_{D-1} + \sum_{n=0}^{D-2} e_n
   \overline{G}_{D-n} \lambda_n + H_3 \Lambda_{D-3} - H_2^A
   \Lambda_{D-2, A}
   \end{equation}
and
  \begin{equation}
  F_{D+1} = d D_D + \sum_{n=0}^{D-2} f_n \overline{G}_{D-n} C_{n+1}
  + H_3 D_{D-2} + H_2^A D_{D-1 ,A}
  \quad ,
  \end{equation}
with $e_n$ and $f_n$ real parameters, and gauge invariance with
respect to $\Sigma$ and $\Lambda$ already implied. Imposing gauge
invariance with respect to $\lambda$ leads to the equations
  \begin{eqnarray}
  & & e_n (-)^{D-n} + f_n =0 \quad \qquad \quad \quad n=0,...,D-2
  \nonumber \\
  & & e_n - f_{n+1} + c_n^{(1)} = 0 \quad \qquad \quad \quad
  n=0,...,D-3  \nonumber \\
  & & -e_n + f_{n+2} (-)^{D-n} + c_n^{(0)} =0 \quad \quad n=0,...,D-4
  \quad .
  \end{eqnarray}
The reader can verify by directly solving the equations that in each
dimension there are two solutions. Given that there are no possible
field redefinitions, as shown in eq.~\eqref{redundancyCC}, this
implies that there are two $D$-form singlets, in agreement with
Table \ref{NSformsanyD}. As an example, we give again the solution
for the five-dimensional case, using the solution
\eqref{3forminfivefixedalpha} and \eqref{4forminfivefixedalpha} for
the coefficients $c_n^{(0)}$ and $c_n^{(1)}$. That is
  \begin{eqnarray}
    & & e_0 =  \beta \qquad e_1 = - \beta + \tfrac{1}{4} \qquad
    e_2 = -\beta +\tfrac{1}{4} \qquad e_3 =\beta \nonumber \\
  & & f_0 =  \beta \qquad f_1 =\beta - \tfrac{1}{4}  \qquad \ \
  f_2 = -\beta + \tfrac{1}{4} \qquad f_3 = -\beta \quad .
  \end{eqnarray}
This solution contains an arbitrary real parameter $\beta$, which is
the relative normalisation of the two $D$-forms.

To conclude, we have determined the gauge transformations of all the
fields in Table \ref{NSformsanyD} for any dimension. In the next
section we will use these results to determine the corresponding WZ
terms, and by analysing the world-volume field content we will show
in which cases we expect these WZ terms to be describing
supersymmetric branes.

\section{String solitons in any dimension}
The aim of this section is to derive a universal expression for the
WZ terms that we claim to describe the charge sector of the
effective action of string solitons in any dimension. This analysis
is the generalisation to any dimension of the one performed in
section 3 in the ten-dimensional case. For any of the forms given in
Table \ref{NSformsanyD}, whose gauge transformations have been
determined in the previous section, one can introduce suitable
world-volume fields that lead to a gauge invariant WZ term. We
expect the world-volume fields to occur in a democratic formulation
together with their electromagnetic duals. We stress that this is a
world-volume duality relating fields in the effective action, which
has nothing to do with the target space electromagnetic duality
relating the various bulk fields in $D$ dimension. After this
electromagnetic duality is taken into account, we expect the
world-volume fields of supersymmetric branes to belong to
half-supersymmetry multiplets. These multiplets are vector
multiplets in all cases, with the exception of the 5-branes, for
which either a vector multiplet or a tensor multiplet can occur.
This will thus be our selection criterion, we select all WZ terms
whose corresponding world volume fields fit within a
half-supersymmetry multiplet.

We first review how the D-brane WZ terms in \cite{Bergshoeff:2010xc}
were constructed. Given that the gauge algebra reviewed in eqs.
\eqref{gaugealgebraalpha=0rev} and \eqref{gaugealgebraalpha=-1rev}
only contains the field strengths of the $\alpha =0$ fields, one
only needs to introduce world-volume fields $b_{0,A}$ and $b_1$
associated to these fields, that is
  \begin{eqnarray}
  & & b_{0,A} \ : \qquad \delta b_{0,A} = - \Sigma_{0,A} \nonumber
  \\
  & & b_1 \ \ \ : \qquad \delta b_1 = -\Sigma_1 \quad ,
  \label{bworldvolumefields}
  \end{eqnarray}
and define the gauge invariant quantities
  \begin{eqnarray}
  & & {\cal F}_{1,A}= d b_{0,A} + B_{1,A} \nonumber \\
  & & {\cal F}_2 = d b_1 + B_2 - H_{2,A} b_0^A \quad .
  \end{eqnarray}
This leads to the gauge-invariant WZ terms in eqs. \eqref{WZstring}
and \eqref{WZterm2}.

It is worth showing how eq. \eqref{WZterm2} leads to the correct
number of worldvolume degrees of freedom for any $p$-dimensional
worldvolume in $D$ dimensions. This analysis was not performed in
full detail in \cite{Bergshoeff:2010xc}. The WZ term \eqref{WZterm2}
contains the worldvolume vector $b_1$ together with $2d$ worldvolume
scalars $b_{0,A}$. The vector corresponds to $p-2$ degrees of
freedom, and considering also the $D-p$ embedding scalars one needs
$d$ additional worldvolume scalars to fill the bosonic sector of a
vector multiplet in $p$ dimensions, while there are $2d$ scalar
fields $b_{0,A}$, that is twice as many. We now show that in eq.
\eqref{WZterm2} only half of the $b_{0,A}$ fields occur. To
understand the mechanism, it is enough to expand \eqref{WZterm2} and
consider only the first ${\cal F}_{1,A}$ term, that is
  \begin{equation}
  C_{p,\alpha} + {\cal F}_{1,A} \Gamma^A_\alpha{}^\beta C_{p-1,
  \beta} + ... \quad .
  \end{equation}
We move to an $\text{SO}(d,d)$ lightcone basis, and we denote the
light-cone directions as $1\pm$, $2\pm$,..., $d\pm$. The analysis of
the properties of the $\text{SO}(d,d)$ Gamma matrices in the
lightcone basis is performed in detail in Appendix A. For a given
value of the spinor index $\alpha$, one can show that for any fixed
$n=1,...,d$ only one of the two matrices $\Gamma_{n\pm,
\alpha}{}^\beta$ gives a non-zero result when acting on a chiral
spinor. This can be seen explicitly in the basis chosen in Appendix
A, see e.g. eqs. \eqref{appendixlightconegammad=1},
\eqref{appendixlightconegammad>22+2-} or
\eqref{appendixlightconegammad>23+} and
\eqref{appendixlightconegammad>23-}, and one can show using the
$\text{SO}(d,d)$ Gamma matrices given in
\eqref{representationClifford} and \eqref{from2dtoddmultiplyi} and
the lightcone ones given in \eqref{lightconegammadef} that this
result is completely general. This analysis proves that for any
given D-brane only half of the $2d$ worldvolume scalars actually
occurs, and this results in the correct number of degrees of freedom
for a $p$-dimensional worldvolume vector multiplet.

We next consider the $D$ fields, whose gauge algebra has been
derived in the previous section. The corresponding gauge
transformations contain both the field strengths of the $B$ and of
the $C$ fields, which implies that in order to write down a gauge
invariant WZ term one has to introduce, together with the
world-volume fields in \eqref{bworldvolumefields}, the world-volume
fields $c_{n, \alpha}$ such that
 \begin{equation}
 \delta c_{n, \alpha} = - \lambda_{n,\alpha} \quad .
 \end{equation}
This implies that these fields are $\text{SO}(d,d)$ spinors of one
chirality for $n$ even and of opposite chirality for $n$ odd,
  \begin{equation}
  c_{2n, \alpha} =
\begin{pmatrix}
c_{2n, a}\\
0
\end{pmatrix} \quad , \qquad c_{2n-1, \alpha} =
\begin{pmatrix}
0\\
c_{2n-1, \dot{a}}
\end{pmatrix} \quad .
\end{equation}
One then introduces the gauge invariant world-volume curvatures
  \begin{equation}
  {\cal G}_{n ,\alpha} = d c_{n-1,\alpha} + C_{n,\alpha} - H_{2,A}
  \Gamma^A_\alpha{}^\beta c_{n-2, \beta} + H_3 c_{n-3 ,\alpha}
  \quad ,
  \end{equation}
whose curl is
  \begin{equation}
  d {\cal G}_{n,\alpha} = G_{n+1 ,\alpha} - H_{2,A}
  \Gamma^A_\alpha{}^\beta {\cal G}_{n-1 ,\beta} - H_3 {\cal G}_{n-2
  , \alpha} \quad . \label{curlofcalG}
  \end{equation}
Using this, one can define a gauge invariant WZ term in all cases.

We first consider the fields $D_{D-4+m ,A_1 ...A_m}$ in
eq.~\eqref{DD-4+mfieldsforanym}, whose gauge transformation is given
in eq.~\eqref{variationDD-4+m}, with the coefficients given in
eqs.~\eqref{valuesofavaluesofb} and \eqref{andbrecursion}. The
corresponding WZ term is given by the universal expression
  \begin{equation}
  \left[ e^{{\cal F}_1} \ \tfrac{m!}{\tilde{m}!} \left( D_{D-4 +
  \tilde{m}}  - \sum_{n=0}^{D-6+\tilde{m}} a_n^{(\tilde{m})}
  (-)^{D+\tilde{m} -n} \overline{\cal G}_{D-5+ \tilde{m}-n}
  \Gamma_{(\tilde{m})} C_{n+1} \right) \right]_{D-4+m, A_1 ...A_m}
  \quad . \label{WZtermuniversalexpression}
  \end{equation}
The rule in this expression is that we have to pick a $D-4+m$ form,
so for any $D-4+ \tilde{m}$-form in the term in curved brackets we
have to pick $m-\tilde{m}$ powers of ${\cal F}_1$. The
$\text{SO}(d,d)$ indices are always meant to be antisymmetrised, and
we have denoted with $\Gamma_{(\tilde{m})}$ the antisymmetric
product of $\tilde{m}$ Gamma matrices. The real coefficients
$a_n^{(\tilde{m})}$ are given in eqs.~\eqref{valuesofavaluesofb} and
\eqref{andbrecursion}. Using eqs.~\eqref{gaugealgebraalpha=-1rev}
and \eqref{variationDD-4+m}, as well as eq.~\eqref{curlofcalG}, one
can show that the variation of eq.~\eqref{WZtermuniversalexpression}
is a total derivative.

The $c_{n,\alpha}$ fields that appear in eq.
\eqref{WZtermuniversalexpression} are $c_{0,\alpha}$,
$c_{1,\alpha}$,..., $c_{D-6+m,\alpha}$. We will impose in all cases
that the $c$ fields are related by electromagnetic duality relations
on the $D-4+m$-dimensional world-volume, which implies that the
field $c_{n,\alpha}$ and the field $c_{D-6+m-n, \alpha}$ are duals.
When $D-4+m$ is even, the field of rank $(D-6+m)/2$ satisfies a
self-duality condition. The reason for assuming this is that we have
already shown in section 3 that this leads to the right counting of
degrees of freedom in ten dimensions. When counting the bosonic
world-volume degrees of freedom, one has also to consider the $4-m$
embedding scalars, together with the extra scalars $b_{0,A}$. We are
counting the degrees of freedom for fixed $A_1...A_m$, where the
indices are all different because of antisymmetry. This implies that
the scalars $b_{0,A}$ contribute as $m$ degrees of freedom.
Therefore,  the embedding scalars plus the scalars $b_{0,A}$
contribute in total as four degrees of freedom in all cases.

We then count the degrees of freedom carried by the $c$ fields. We
recall that the $\text{SO}(d,d)$ chiral spinors $c_{n, \alpha}$ have
$2^{d-1}$ real components. As we will show in a case by case
analysis in the following, one needs  $2^{d-m-1}$ chiral components
to get the right counting of the degrees of freedom in each case.
When $m=1$ the different light-cone $\text{SO}(d,d)$ directions
correspond to independent branes. This is because, as shown in
Appendix A, the light-cone Gamma matrices have the property of
projecting out half of the components of a spinor for each
chirality. As already discussed, we denote the light-cone directions
as $1\pm, 2\pm ,...,d\pm$. For $m=2$, $m=3$ and $m=4$ the request
that the degrees of freedom collect in a multiplet, that is the
request that the product of $m$ Gamma matrices projects a chiral
spinor to $2^{d-m-1}$ chiral components, imposes that one has to
consider only the components corresponding to indices $n_1\pm n_2\pm
... n_m\pm$, with the $n_i$ all different. In formula, we have

\begin{equation}\label{informula}
\Gamma_{(m)}\ \ \rightarrow\ \ \Gamma_{n_1\pm n_2\pm ...
n_m\pm}\,,\hskip 1truecm \text{all}\  n_i \ \text{different}\,.
\end{equation}
As discussed in Appendix A, this forms a conjugacy class, and the
corresponding products of $m$ Gamma matrices satisfy the property of
being nilpotent. In $D=9,...,6$ one has among the rest the case
$m=d$, which is special because in this case the Gamma matrices with
indices $n_1\pm .... n_d\pm$ ($n$'s all different) map a chiral
spinor to either zero or one component (according to the chirality
of the spinor). Given these rules, the outcome of our analysis will
be that eq. \eqref{WZtermuniversalexpression} leads to a
world-volume vector multiplet for $m<d$, and to either a vector or a
tensor multiplet for $m=d$ (which corresponds to a 5-brane),
depending on whether one takes the self-dual or the anti-self-dual
part of the completely antisymmetric $\text{SO}(d,d)$ representation
with $d$ indices. We will also see that one cannot obtain the right
degrees of freedom for $m>d$, so we expect this case not to
correspond to a brane. This result is summarised in Table
\ref{result}. The branes are along the light-cone directions
described above, and their number for each dimension and for each $m
$ is given in Table \ref{conjugacyclasses} (see also Appendix A for
the details).

We now analyse the explicit form of eq.
\eqref{WZtermuniversalexpression} in all the specific cases, that is
for all the different values of $m$ from 0 to 4. The fact that the
counting works for the case $m=0$ actually already guarantees that
it has to work for higher $m$. For instance, one gets for $m=1$ in
$D$ dimensions exactly the same counting as for $m=0$ in $D+1$
dimensions. The reason is that when going from $D+1$ to $D$
dimensions, the number of spinor components for each $c$ field
doubles, which is compensated by the fact that each light-cone Gamma
matrix projects out half of the components. This gives the same
counting for a $D-3$-dimensional world-volume. The same reasoning
applies to the other values of $m$.

\begin{table}[t]
\begin{center}
\begin{tabular}{|c||c|c||c|c||c|c|}
\hline
 \rule[-1mm]{0mm}{6mm} $D$ & \multicolumn{2}{c||}{$m=2$} & \multicolumn{2}{c||}{$m=3$} & \multicolumn{2}{c|}{$m=4$}
 \\
  \hline
  \hline
 \rule[-1mm]{0mm}{6mm} 8 & $ {\bf 3}^+ + {\bf 3}^-$ & $2+2$ &  & &&
 \\
 \hline
 \rule[-1mm]{0mm}{6mm} 7 & $ {\bf 15}$ & 12 & ${\bf 10}^+ + {\bf 10}^-$ & $4+4$
 & &
 \\
\hline
 \rule[-1mm]{0mm}{6mm} 6 & $ {\bf 28}$ & 24 & ${\bf 56}$ & 32 &
 ${\bf  35}^+ + {\bf 35}^-$ & $8+8$ \\
 \hline
  \rule[-1mm]{0mm}{6mm} 5 & $ {\bf 45}$ & 40 & ${\bf 120}$ & 80 &
 ${\bf 210}$ & $80$ \\
 \hline
\end{tabular}
\end{center}
\caption{\sl Table giving the dimension of the conjugacy classes of
the solitonic branes in various dimensions. For each $m$, we denote
in the first column the $\text{SO}(d,d)$ representation of the field
$D_{D-4+m, A_1 ...A_m}$, whose dimension is ${2d \choose m}$, and in
the second column the dimension of the conjugacy class, $2^m {d
\choose m}$. For $d=m$ the representation splits in selfdual and
anti-selfdual, corresponding to branes supporting either a vector or
a tensor multiplet. }
  \label{conjugacyclasses}
\end{table}

\subsection*{$m=0$}
For $m=0$, eq. \eqref{WZtermuniversalexpression} collapses to the
expression
  \begin{equation}
  D_{D-4} - \sum_{n=0}^{D-6} a_n^{(0)}
  (-)^{D -n} \overline{\cal G}_{D-5-n} C_{n+1}
  \label{WZtermofD-5brane}
  \end{equation}
containing the world volume fields
$c_{0,\alpha}$,...,$c_{D-6,\alpha}$, with $c_{n,\alpha}$ dual to
$c_{D-6-n,\alpha}$. In addition to these fields, one has to consider
four embedding scalars in any dimension. We now analyse this in
different dimensions.

The ten-dimensional case has already been discussed in section 3. We
have shown that in the IIA case the fields are $c_0$, $c_2$ and
$c_4$. Imposing duality, one obtains one scalar plus one self-dual
tensor, that together with the four embedding scalars makes the
bosonic field content of a tensor multiplet in six dimensions. This
is the WZ term of an NS5A brane. In the IIB case, instead, one has
$c_1$ and $c_3$, which corresponds to a vector, and together with
the four embedding scalars one obtains a vector multiplet, which is
the WZ term of an NS5B brane.

In nine dimensions one has a five-dimensional world volume. The
world-volume field content is $c_0$, $c_1$, $c_2$ and $c_3$ (there
are no indices because the chiral spinors of $\text{SO}(1,1)$ are
one-dimensional). Imposing duality, one obtains a vector and a
scalar, that together with the four embedding scalars makes a vector
multiplet in five dimensions. We thus expect this to be the WZ term
of a 4-brane.

In eight dimensions the world-volume is four-dimensional, and the
fields are $c_{0,a}$, $c_{1,\dot{a}}$ and $c_{2,a}$, two-dimensional
spinors of the T-duality group $\text{SO}(2,2)$. After duality, this
corresponds to two scalars and one vector, which together with the
four embedding scalars makes a vector multiplet in four dimensions.
We consider this the WZ term of a 3-brane in eight dimensions.

In seven dimensions one has a three-dimensional world-volume, with
fields $c_{0,a}$ and $c_{1,\dot{a}}$, $a,\dot{a}=1,...,4$. These
fields are dual in three dimensions, and together with the four
embedding scalars this makes the eight bosonic degrees of freedom of
a three-dimensional scalar multiplet. Similarly, in six dimensions
one has a two-dimensional 8-component scalar $c_{0,a}$ which
satisfies a self-duality condition, that is four degrees of freedom.
Again, together with the embedding scalars this sums up to eight
bosonic degrees of freedom.

To summarise, we thus expect in all cases the WZ term of eq.
\eqref{WZtermofD-5brane} to correspond to a supersymmetric
$D-5$-brane.

\subsection*{$m=1$}
Eq. \eqref{WZtermuniversalexpression}, evaluated for $m=1$, gives
  \begin{eqnarray}
   & & D_{D-3,A} + \sum_{n=0}^{D-5} a_n^{(1)}
  (-)^{D -n} \overline{\cal G}_{D-4-n} \Gamma_A C_{n+1}\nonumber \\
  & &   + {\cal F}_{1,A} \left(
  D_{D-4} - \sum_{n=0}^{D-6} a_n^{(0)}
  (-)^{D -n} \overline{\cal G}_{D-5-n} C_{n+1} \right) \quad .
  \label{WZtermofD-3brane}
  \end{eqnarray}
Clearly, this expression only makes sense in dimension lower than
ten. To count the expected degrees of freedom, one has to split the
$A$ index in light-cone coordinates, and each light-cone coordinate
corresponds to $2^{d-2}$ components for the chiral spinor.

In nine dimensions (that is for $d=1$) each of the two light-cone
coordinates project on either one chirality or the opposite. Indeed,
for $\text{SO}(1,1)$ splitting $\Gamma_A$ in light-cone coordinates
corresponds to taking the self-dual or the anti-self-dual
combination (see Appendix A). This means that in the anti-selfdual
case one gets $c_0$, $c_2$ and $c_4$, and in the selfdual case one
gets $c_1$ and $c_3$, in both cases together with 4 additional
scalars. Imposing duality one gets a 5-brane supporting a tensor
multiplet in the former case and a vector multiplet in the latter.

In eight dimensions each light-cone component selects a single
spinor component out of the two components for each chirality. This
leaves the fields $c_0$, $c_1$, $c_2$ and $c_3$ in a
five-dimensional world-volume.  Imposing duality, this leads to a
vector multiplet for each light-cone component. This shows how the
halving of the number of spinor components with respect to the $m=0$
case is crucial to get the right counting.

In seven dimensions each light-cone Gamma matrix projects on two
components out of the four spinor components for each chirality.
This gives two $c_0$'s, two $c_1$'s and two $c_2$'s. Imposing
self-duality on the four-dimensional world-volume one gets a vector
and two scalars, that together with the additional four scalars
makes the bosonic sector of a vector multiplet in four dimensions.

The same applies to the lower dimensional case. In general, as
already stressed, in each dimension $D$ the counting of the degrees
of freedom of the $m=1$ branes is the same as the counting of the
degrees of freedom of the $m=0$ branes in $D+1$ dimensions.

\subsection*{$m=2$}
In this case in nine dimensions one would expect a 6-brane, that is
a 7-dimensional world-volume, but there is no way of selecting the
fields of a seven-dimensional vector multiplet, because there is no
way of getting rid of the world-volume fields $c_2$ and $c_3$, which
would correspond to a seven-dimensional rank-2 tensor. We thus
expect this not to correspond to a brane. This applies to all the
other cases $m>d$.

In eight dimensions, we take the light-cone directions $1\pm 2\pm$.
For each of these four components one selects out of a 2-component
chiral spinor either one component or zero, according to the
chirality. The directions $1+2+$ and $1-2-$ select $c_1$ and $c_3$,
while the directions $1+2-$ and $1-2+$ select $c_0$, $c_2$ and
$c_4$. In the first case one gets a vector multiplet, and in the
second a tensor multiplet. Therefore, there are two vector branes
and two tensor branes, as shown in Table \ref{conjugacyclasses}.

In seven dimensions, out of the 15 $\text{SO}(d,d)$ components of
$D_{5,AB}$, one selects the 12 components $1\pm2\pm$, $1\pm 3\pm$
and $2\pm 3\pm$. The products of Gamma matrices in these directions
select a single components out of the four-component spinors
$c_{n,\alpha}$, resulting in a vector multiplet in a
five-dimensional world-volume. All the lower-dimensional cases are
analogous, one obtains in $D$ dimensions the same counting of the
case $m=1$ in $D+1$ dimensions.

\subsection*{$m=3$}
This case does not exist in nine dimensions, while in eight
dimensions it does not lead to a vector multiplet, exactly like the
$m=2$ case in nine dimensions. In seven dimensions one considers the
directions $1\pm 2\pm 3\pm$, which is eight components. Four of the
corresponding Gamma matrices project on a single component of one
chirality, and the other four of a single component of the opposite
chirality. This gives a six-dimensional world-volume with either a
vector or a tensor multiplet.

In lower dimensions $D$ one gets $(D-2)$-branes supporting vector
multiplets. These branes correspond to the light-like directions
$n_1\pm n_2 \pm n_3 \pm$, with $n_1$, $n_2$ and $n_3$ all different,
and their number is given in Table \ref{conjugacyclasses}. Again,
the counting is precisely the same as the one of the previous case
$m=2$ in $D+1$ dimensions.

\subsection*{$m=4$}
In this case the branes are spacetime-filling. One does not get a
vector multiplet in eight and seven dimensions, while in six
dimensions one takes the 16 components $1\pm2\pm3\pm 4\pm$. The
corresponding products of Gamma matrices project on a single
component of a given chirality. This gives 8 branes supporting a
vector multiplet and 8 branes supporting a tensor multiplet.

In lower dimensions $D$ one gets $(D-1)$-branes supporting vector
multiplets. These branes correspond to the light-like directions
$n_1\pm n_2 \pm n_3 \pm n_4 \pm$, with $n_1$, $n_2$, $n_3$ and $n_4$
all different, and their number is given in Table
\ref{conjugacyclasses}. Again, as for all the other cases the
counting is precisely the same as the one of the previous case $m=3$
in $D+1$ dimensions.

\vskip 1cm

\noindent We finally comment on the possibility that the other
fields in Table \ref{NSformsanyD} lead to WZ terms that satisfy our
criteria. One simple reason why this is not the case is that these
WZ term must contain the field strength ${\cal F}_2$, because the
corresponding fields contain $H_3$ in their field strengths and
gauge transformations, differently from the fields $D_{D-4+m ,A_1
...A_m}$ in eq. \eqref{DD-4+mfieldsforanym} (see section 4). This
would thus lead to an additional world-volume vector, which would
make it impossible to form a world-volume multiplet.

In principle one should also consider the possibility of taking
linear combinations of fields in Table \ref{NSformsanyD} that belong
to the same representation. This is the case for the two 7-forms in
9 dimensions, that are both $\text{SO}(1,1)$ singlets, for the two
7-forms in 8 dimensions, that are both $\text{SO}(2,2)$ vectors, and
of (some of) the $D$-forms in nine, eight and seven dimensions. We
have investigated this possibility, and the outcome of our analysis
is that there is no possible linear combination that selects the
right degrees of freedom.

To summarise, none of these additional fields give WZ terms that
satisfy our criteria, and therefore our final result is the one
summarised in Tables \ref{result} and \ref{conjugacyclasses}.

\section{Central charges}

It is interesting to compare our results with what we know about the
central charges of the target space Poincare supersymmetry algebra.
These  charges are $n$-forms that transform as representations of
the $R$-symmetry of the supersymmetry algebra, see Table
\ref{centralcharges}. In general an $n$-form central charge relates
to a $p$-brane with $p$ or $p=D-n$. The relation between central
charges and branes \cite{Townsend:1997wg} only applies to
asymptotically flat branes and, therefore, can only be applied to
the solitonic $D_{D-4}$ and $D_{D-3,A}$ fields but not to the
higher-form $D$-fields. We therefore only consider $(D-5)$-branes
and $(D-4)$-branes implying that we consider only solitonic branes
for $D\ge 4$. These particular branes, corresponding to $m=0$ and
$m=1$,  do not transform in non-trivial conjugacy classes. Below we
discuss the different dimensions separately and show how all central
charges correspond to the branes we have been discussing in this
paper.
\bigskip

\begin{table}[t]
\begin{center}
\begin{tabular}{|c|c|c|c|c|c|c|c|}
\hline
$D$&$R$-symmetry&$n=0$&$n=1$&$n=2$&$n=3$&$n=4$&$n=5$\\[.1truecm]
\hline \rule[-1mm]{0mm}{6mm} IIA&{\bf 1}&{\bf 1}&{\bf 1}&{\bf 1}&--&{\it 1}&${\it 1}$\\[.05truecm]
\hline \rule[-1mm]{0mm}{6mm} IIB&SO(2)&--&{\bf 2}&--&{\bf 1}&--&${\it 1}^+ + {\it 2}^+$\\[.05truecm]
\hline \rule[-1mm]{0mm}{6mm} 9&SO(2)&${\bf 1}+{\bf 2}$&{\bf 2}&{\bf 1}&{\bf 1}&${\it 1}+{\it 2}$&\\[.05truecm]
\hline \rule[-1mm]{0mm}{6mm} 8&U(2)& $ 2 \times {\bf 3}$ &{\bf 3} &
$2 \times {\bf 1}$& ${\it 1} + {\it 3}$
&${\it 3}^+ + {\it 3}^-$&\\[.05truecm]
\hline \rule[-1mm]{0mm}{6mm} 7&Sp(4)& {\bf 10} & {\bf 5} & ${\it 1} +{\it 5}$ &{\it 10} &&\\[.05truecm]
\hline \rule[-1mm]{0mm}{6mm} 6&Sp(4)$\times$Sp(4)&$({\bf 4},{\bf 4})$&$({\it 1},{\it 1})$ & $({\it 4},{\it 4})$& $({\bf 10},{\bf 1})^+$ &&\\[.05truecm]
\rule[-1mm]{0mm}{6mm} & & &$({\it 1},
{\it 5})$ & &$({\bf 1},{\bf 10})^-$ &&\\[.05truecm]
\rule[-1mm]{0mm}{6mm} & & &$({\it 5},
{\it 1})$ & &&&\\[.05truecm]
\hline \rule[-1mm]{0mm}{6mm} 5&Sp(8)& {\it 1} + {\it 27}& {\it 27}& {\bf 36}&&&\\[.05truecm]
\hline \rule[-1mm]{0mm}{6mm} 4&SU(8)& ${\it 28} + {\it \overline{28}}$ & {\bf 63} & ${\bf 36}^+ + {\bf \overline{36}}^-$ &&&\\[.05truecm]
\hline \rule[-1mm]{0mm}{6mm} 3&SO(16)& {\bf 120} & {\bf 135} &&&&\\[.05truecm]
\hline
\end{tabular}
\end{center}
  \caption{\sl This table indicates the $R$-representations of the
  $n$-form central charges of $3\le D\le 10$ maximal supergravity.
  If applicable, we have also indicated the space-time duality of
  the central charges with a superscript $\pm$. The central charges
  that are discussed in the text are indicated in italic. }
  \label{centralcharges}
\end{table}

%eric
\noindent ${\bf IIA}$\ \ In the ten-dimensional IIA theory  we
consider 5-branes and 6-branes which are described by  the 5-form
and the dual of the 4-form central charges, respectively. We first
consider the 6-branes. The IIA theory contains only a Dirichlet
6-brane with a vector multiplet. This D6-brane is described by (the
dual of) the ${\bf 1}$ 4-form central charge. We next consider the
5-branes which are described by the 5-form central charges. In the
D-brane sector there are no 5-branes. In the solitonic sector there
is a solitonic 5-brane with a tensor multiplet. Finally, in the
gravitational sector there is a KK monopole with a vector multiplet
\cite{Bergshoeff:1997gy}. In total we find two 5-branes, one with a
vector and one with a tensor multiplet. These two branes are
described by the ${\bf 1}$ 5-form central charge and its dual.
\bigskip

\noindent ${\bf IIB}$\ \ Like in the IIA case, we consider 5-branes
and 6-branes which are described by  the 5-form and the dual of the
4-form central charges, respectively. The IIB theory has no 6-branes
and, indeed there is no 4-form central charge. We next consider the
5-branes. In the D-brane sector the IIB theory contains a D5-brane
with a vector multiplet. In the solitonic sector it contains a
solitonic 5-brane with a  vector multiplet and in the gravitational
sector it contains a KK monopole with tensor multiplet
\cite{Bergshoeff:1997gy}. In total we find three 5-branes.  The KK
monopole, with the tensor multiplet, is described by the singlet
${\bf 1}^+$ self-dual central charge whereas the D5-brane and the
solitonic 5-brane, with the vector multiplets, are described by the
S-doublet ${\bf 2}^+$ of self-dual central charges.
\bigskip

\noindent ${\bf D=9}$\ \ In $D=9$ dimensions we consider the
4-branes and 5-branes which are both described by the 4-form central
charges. We first consider the 5-branes. In the D-brane sector there
is a Dirichlet 5-brane with a vector multiplet that transforms as a
one-dimensional chiral spinor representation of $\text{SO}(1,1)$. In
the soliton sector there is a 2-dimensional T-duality vector of
solitons that splits into a tensor soliton and a vector soliton. The
tensor soliton corresponds to the ${\bf 1}$ dual 4-form central
charge and the Dirichlet 5-brane together with the vector soliton
correspond to the ${\bf 2}$ dual 4-form central charge. In the
4-brane case we have in the D-brane sector a Dirichlet 4-brane with
a vector multiplet that is a one-dimensional chiral spinor of
$\text{SO}(1,1)$ and in the soliton sector a singlet vector soliton.
Furthermore, in the gravitational sector we have a KK monopole with
a vector multiplet. The KK monopole corresponds to the ${\bf 1}$
4-form central charge and the Dirichlet 4-brane together with the
vector soliton correspond to the ${\bf 2}$ 4-form central charge.
\bigskip

\noindent ${\bf D=8}$\ \ In $D=8$ dimensions we consider the
3-branes and 4-branes which are described by the 3-form and 4-form
central charges. From now on all branes have vector multiplets. We
first consider the 4-branes. In the D-brane sector there is a
two-dimensional T-duality spinor of Dirichlet 4-branes. In the
soliton sector there is a 4-dimensional T-duality vector of
solitons. In total this gives 6 4-branes corresponding to the 6
${\bf 3}^+ + {\bf 3}^-$ 4-form central charges where each Dirichlet
4-brane together with 2 solitonic branes form a triplet. We next
consider the 3-branes. In the D-brane sector there is a 2-component
T-duality spinor of Dirichlet 3-branes. In the soliton sector there
is a singlet solitonic 3-brane and in the gravitational sector there
is a single KK monopole. This adds up to 4 branes corresponding to
the 4 3-form central charges. The KK monopole corresponds to the
${\bf 1}$ central charge while the 2 Dirichlet branes together with
the solitonic brane correspond to the ${\bf 3}$ central charges.
\bigskip

\noindent ${\bf D=7}$\ \ In $D=7$ dimensions we consider 2-branes
and 3-branes corresponding to 2-form and 3-form central charges. We
first consider the 3-branes. From the D-brane sector we have a
4-component T-duality spinor of Dirichlet 3-branes. In the solitonic
sector we have a 6-component T-duality vector of solitons. Together,
they correspond to the ${\bf 10}$ 3-form central charges. We next
consider the 2-branes. In the D-brane sector we have a 4-component
T-duality spinor of Dirichlet 2-branes while we have a singlet
solitonic 2-brane and KK monopole. The KK monopole corresponds to
the ${\bf 1}$ 2-form central charge while the 4 Dirichlet 2-branes
and singlet soliton correspond to the ${\bf 5}$ 2-form central
charge.
\bigskip

\noindent $\bf D=6$\ \ In $D=6$ dimensions we consider 1-branes and
2-branes corresponding to the 1-form and 2-form central charges. We
first consider the 2-branes. We have an 8-component T-duality spinor
of Dirichlet 2-branes and an 8-component T-duality vector of string
solitons. Together, they correspond to the 16 $({\bf 4},{\bf 4})$
2-form central charges. We next consider the 1-branes. In this case
also the fundamental sector contributes with a singlet fundamental
string. On top of that we have an 8-component T-duality spinor of
Dirichlet 1-branes, a singlet string soliton and a singlet KK
monopole adding up to a total of 11 1-branes. The KK monopole
corresponds to the  $({\bf 1},{\bf 1})$ 1-form central charge. Four
Dirichlet branes together with the fundamental string correspond to
the $({\bf 5},{\bf 1})$ 1-form central charges while  the remaining
4 Dirichlet branes together with the string soliton correspond to
the $({\bf 1},{\bf 5})$ 1-form central charges.
\bigskip

\noindent ${\bf D=5}$\ \ In $D=5$ dimensions we consider 0-branes
and 1-branes corresponding to the 0-form and 1-form central charges.
We first consider the 1-branes. We have a singlet fundamental
string, a 16-component T-duality spinor of Dirichlet 1-branes and a
10-component T-duality vector of string solitons. Together, these 27
1-branes correspond to the ${\bf 27}$ 1-form central charges. We
next consider the 0-branes. We have a 10-component T-duality vector
of fundamental 0-branes, a 16 component T-duality spinor of
Dirichlet 0-branes, a singlet string soliton and a singlet KK
monopole adding up to a total of 28 branes. The KK monopole
correspond to the ${\bf 1}$ 0-form central charge while the
remaining 27 0-branes correspond to the ${\bf 27}$ 0-form central
charge.
\bigskip

\noindent ${\bf D=4}$\ \ Finally, we consider $D=4$ dimensions with
0-branes only. We have a 12-component T-duality vector of
fundamental 0-branes, a 32-component T-duality spinor of Dirichlet
0-branes and a 12-component T-duality vector of solitonic 0-branes
adding up to a total of 56 branes. Sixteen of the Dirichlet 0-branes
together with the 12 fundamental 0-branes correspond to the ${\bf
28}$ 0-form central charges while the remaining 16 Dirichlet
0-branes together with the 12 solitonic 0-branes correspond to the
other ${\bf 28}$ 0-form central charges. \vskip .2truecm

We conclude that the T-duality properties of the fundamental branes,
Dirichlet branes, solitonic branes and KK monopoles precisely fit
with the counting of the central charges in the supersymmetry
algebra. This nicely confirms the results obtained in this paper.

\section{$\text{E}_{11}$ and a ten-dimensional origin}

In section 2 we have derived the T-duality representations of all solitonic
$p$-form potentials in $D<10$ dimensions by decomposing the known U-duality representations into
T-duality representations. Remarkably, unlike D-branes, these $D<10$ T-duality representations cannot be obtained by a
toroidal reduction of the $D=10$ solitonic $p$-form potentials.
In other words, the lower
dimensional $p$-form potentials  do not possess a ten-dimensional origin in the context of the established IIA and IIB supergravity multiplets.
One needs further $D=10$ representations to obtain the complete
$D<10$ T-duality multiplets. One way to achieve this, is to assume that
additional mixed-symmetry representations can be added to the ten-dimensional supergravity
multiplets without upsetting the counting of degrees of freedom.
It turns out that precisely those mixed-symmetry fields are needed that are
predicted by the Kac-Moody algebra $\text{E}_{11}$. With the present technology, it is not known
how to extend the existing supergravity multiplets with mixed-symmetry fields at the level of the full
non-linear supersymmetry algebra. This is only understood at the linearized level. Nevertheless, it is of interest
to see that, once we assume that mixed-symmetry fields can be included,  the $\text{E}_{11}$ Kac-Moody
algebra predicts precisely which fields are needed to re-obtain T-duality after toroidal compactification.

With the above motivation in mind we will show in this section how an $\text{E}_{11}$ analysis
can be used not only to derive  all solitonic $p$-form potentials of IIA and IIB supergravity but also
to derive all
solitonic mixed symmetry fields that occur in the IIA and IIB $\text{E}_{11}$  spectrum. In the first part of this section we will
determine all solitonic fields, be it $p$-form potentials or mixed-symmetry fields, that arise in the
$\text{E}_{11}$ decomposition relevant for the IIA and the IIB
theory. Once these fields have been determined, we will show in the second part of this section how, after a toroidal reduction, they give rise to precisely the same $D<10$ solitonic T-duality multiplets derived earlier in section 2
by decomposing
in each dimension the U-duality representations of each $p$-form with
respect to T-duality, with the result  summarised in Table
\ref{NSformsanyD}.\footnote{The strategy here is similar to the one in \cite{Riccioni:2007au}, where
all the eleven-dimensional $\text{E}_{11}$ fields giving rise to
forms after dimensional reduction down to any dimension above two
were determined.}

%The aim of this section is to re-derive all the $\alpha =-2$ forms
%of maximal supergravity theories in any dimension using the
%ten-dimensional IIA and IIB decompositions of the Kac-Moody algebra
%$\text{E}_{11}$. We will perform in the first part of the section an
%$\text{E}_{11}$ analysis to determine all the fields arising in the
%$\text{E}_{11}$ decomposition relevant for the IIA and the IIB
%theory that give rise to $\alpha =-2$ forms after dimensional
%reduction. Most of these fields have mixed-symmetry indices, and are
%not proper fields of the IIA or IIB theory, in the sense that one
%can not close the supersymmetry algebra on them (this can only be
%done at the linearised level). Still, the supersymmetry algebra
%closes on the dimensionally reduced fields as long as these are
%forms in the non-compact directions. In a sense, the occurrence of
%these fields signals the fact that the corresponding lower
%dimensional forms do not possess a ten-dimensional origin.

%In the second part of the section we will then use
%these results to explicitly determine from dimensional reduction the
%solitonic (that is $\alpha =-2 $) forms in any dimension.
%The result
%clearly coincides with the one obtained in section 2 by decomposing
%in each dimension the U-duality representations of each form with
%respect to T-duality, and it is summarised in Table
%\ref{NSformsanyD}.

\subsection{$\text{E}_{11}$ and mixed-symmetry fields}

In this subsection we will construct a procedure that selects out of the
IIA and IIB $\text{E}_{11}$ spectrum precisely those $p$-forms and mixed-symmetry fields that are solitonic, i.e.~correspond to $\alpha=-2$.
For the reader who wants to skip the technical derivation the final result, which coincides for the IIA and IIB theory,
can be found in eq.~\eqref{m=2fieldsIIBanyl}.

%As shown in \cite{Riccioni:2007au,Bergshoeff:2007qi}, all the forms
%on which the supersymmetry algebra of maximal supergravity theories
%closes in any dimensions below 10 are derived from the Kac-Moody
%algebra $\text{E}_{11}$. These fields include not only all the
%propagating forms and their electromagnetic duals, but also the
%$D-1$-forms and the $D$-forms, which are not propagating but have a
%crucial role in both determining all possible gauged/massive
%deformations of the theory and in being associated to the branes of
%codimension 1 and 0 that are present.

\begin{figure}[h]
\begin{center}
\begin{picture}(380,60)
\multiput(10,10)(40,0){10}{\circle{10}}
\multiput(15,10)(40,0){9}{\line(1,0){30}} \put(290,50){\circle{10}}
\put(290,15){\line(0,1){30}} \put(8,-8){$1$} \put(48,-8){$2$}
\put(88,-8){$3$} \put(128,-8){$4$} \put(168,-8){$5$}
\put(208,-8){$6$} \put(248,-8){$7$} \put(288,-8){$8$}
\put(328,-8){$9$} \put(365,-8){$10$} \put(300,47){$11$}
\end{picture}
\caption{\sl The $\text{E}_{11}$ Dynkin diagram.\label{E11Dynkindiagram}}
\end{center}
\end{figure}
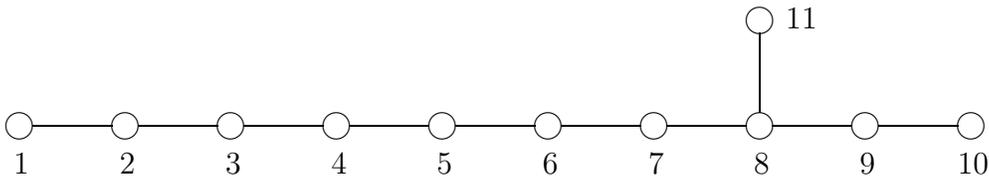

Given the Dynkin diagram of
$\text{E}_{11}$ in Fig. \ref{E11Dynkindiagram}, deleting any node in
the horizontal line other than node 10 (in the nine-dimensional case
one has to delete both nodes 9 and 11) results in a subgroup
$\text{GL}(D,\mathbb{R}) \times \text{E}_{11-D}$, and the positive
level generators\footnote{Given the decomposition of a positive root
in terms of simple roots, the ``level'' associated to a given simple
root is the number of times this simple root occurs.} with
completely antisymmetric $\text{GL}(D,\mathbb{R})$ indices are
associated to the fields of the $D$ dimensional supergravity theory.
In particular, in \cite{Riccioni:2007au} this decomposition was
obtained by explicitly performing a dimensional reduction of the
generators associated to eleven dimensions, that is the positive
level generators that are representations of the
$\text{GL}(11,\mathbb{R})$ subgroup that results from deleting node
11. Most of these generators do not belong to completely
antisymmetric representations, and therefore the corresponding
fields are not forms. Nonetheless, some of these fields give rise to
forms after dimensional reduction. Correspondingly, the
supersymmetry algebra does not close on these fields in the eleven
dimensional theory (it does only at the linearised level), but it
does in the lower dimensional one as long as in the lower
dimensional theory they are forms.

In \cite{West:2001as} eleven dimensional supergravity was originally
shown to be associated to the $\text{E}_{11}$ algebra after a
decomposition that results from deleting node 11, while its
dimensional reduction to the IIA theory is obtained by considering
the decomposition that results from the further deletion of node 10
\cite{West:2001as}. The IIB theory was shown to result from
$\text{E}_{11}$ decomposed via the deletion of node 9
\cite{Schnakenburg:2001ya}, while the full form field content of
both IIA and IIB, including the 10-forms, was obtained in
\cite{Kleinschmidt:2003mf}. This was then shown
\cite{Bergshoeff:2005ac} to be the field content on which the IIA
and IIB supersymmetry algebras close.

In this paper we want to determine how the fields of
lower-dimensional maximal supergravities form representations of the
T-duality group, and we are thus interested in the dilaton weight of
the ten-dimensional $\text{E}_{11}$ fields. We consider the
decomposition of the $\text{E}_{11}$ algebra which is appropriate to
either the IIB or the IIA theory. As already recalled, the IIB and
the IIA generators correspond to decomposing the $\text{E}_{11}$
algebra with respect to the two $\text{A}_9$ subalgebras that arise
from either deleting node 9 (IIB case) or nodes 10 and 11 (IIA
case). In both cases, the way the fields scale with respect to the
dilaton is dictated by the level associated to node 10, which we
call $m$ \cite{Cook:2008bi}. This means that in the corresponding
positive root of $\text{E}_{11}$ the simple root $\alpha_{10}$
occurs $m$ times. The relation between $m$ and $\alpha$ is simply
\cite{Cook:2008bi}\footnote{The parameter $\alpha$ should not be
confused with the $\text{E}_{11}$ roots, which are also denoted by
$\alpha$.}
  \begin{equation}
  m =-\alpha
  \quad .
  \end{equation}

We use now the strategy of refs.
\cite{Gaberdiel:2002db,Damour:2002cu,West:2002jj} to decompose the
roots of $\text{E}_{11}$ level by level in terms of representations
of a finite dimensional subalgebra. We first concentrate on the IIB
case. The decomposition of the $\text{E}_{11}$ roots which is
appropriate to IIB was performed in \cite{Riccioni:2006az}, and here
we review these results. We denote with $\alpha_i$ the simple roots
associated to nodes 1-8 and node 11. These are the simple roots of
the $\text{A}_9$ of IIB. Given a positive root $\alpha$ of
$\text{E}_{11}$, one can then write
  \begin{equation}
  \alpha = \sum_i n_i \alpha_i + l \alpha_9 + m \alpha_{10} \quad ,
  \end{equation}
where $l$ and $m$ are the levels associated to nodes 9 and 10. One
then writes the simple root $\alpha_9$ as
  \begin{equation}
  \alpha_9 = x - \frac{1}{2} \alpha_{10} - \lambda_8 \quad ,
  \end{equation}
where $x$ is a vector in the direction orthogonal to the simple
roots $\alpha_i$ and $\alpha_{10}$, and where $\lambda_8 $ is the
fundamental weight of the $\text{A}_9$ subalgebra associated to node
8, that is
  \begin{equation}
  \lambda_8  = \frac{1}{5} \left[ \alpha_1 + 2 \alpha_2 +3\alpha_3 +
  ...+ 8 \alpha_8 + 4 \alpha_{11} \right] \quad ,
  \end{equation}
and
  \begin{equation}
  \lambda_8^2 = \frac{8}{5} \quad .
  \end{equation}
This gives
 \begin{equation}
 x^2 = - \frac{1}{10} \quad , \label{xsquaredinIIB}
 \end{equation}
given that all the simple roots have square length 2. In order to
have a representation of $\text{A}_9$ denoted by the Dynkin indices
$p_i$ in the adjoint of $\text{E}_{11}$, a necessary condition is
that the highest weight $\sum_i p_i \lambda_i $ occurs when one
projects the positive roots along the space of the simple roots
$\alpha_i$ of $\text{A}_9$. This means that one must have
  \begin{equation}
  \alpha = \sum_i p_i \lambda_i + l x + \left(m - \frac{l}{2} \right)
  \alpha_{10} \quad .
  \end{equation}
As shown in \cite{Riccioni:2006az}, there is a relation between the
level $l$ and the number of $\text{GL}(10,\mathbb{R})$ indices of
the corresponding generators. This relation is dictated by the fact
that the root $\alpha_{9}$ corresponds to a generator with 2
antisymmetric indices, and thus at level $l$ one finds that all
generators must have $2l$ indices. Including also the possibility
that there are groups of 10 antisymmetric indices, one obtains the
condition
  \begin{equation}
  10 n + \sum_i p_i (10-i ) = 2 l \quad ,
  \end{equation}
where $n$ is the number of groups of 10 antisymmetric indices.
Another constraint comes from imposing that the roots must have
square length at most 2, or more precisely $\alpha^2 = 2, 0 ,
-2,...$ \cite{Kac}. Using eq. \eqref{xsquaredinIIB} one
gets\footnote{This formula is identical to eq. (27) of ref.
\cite{Riccioni:2006az}, as can be seen using eq. (26) in that paper,
which relates the level $m$ to the $\text{SL}(2,\mathbb{R})$ weight
$q$.}
 \begin{equation}
 \alpha^2 = \sum_{ij} p_i A_{ij}^{-1} p_j + F_B (l,m) \quad ,
 \end{equation}
where
 \begin{equation}
 F_B (l,m) = \frac{2}{5} l^2 + 2 m^2 -2ml \label{FBIIB}
 \end{equation}
 and  where $A_{ij}^{-1}$ is the inverse of the $\text{A}_9$ Cartan matrix,
  \begin{equation}
  A_{ij}^{-1} = (\lambda_i , \lambda_j )\,.
  \end{equation}
 For the general  $\text{A}_{n-1}$ algebra the inverse matrix is given by
  \begin{equation}
  (A_{jk})^{-1}=
  \begin{cases}{j(n-k)\over n}, \ \ j\le k\\
  {k(n-j)\over n}, \ \ j\ge k
  \end{cases}
  \quad .
  \end{equation}

The strategy is now to analyse all possible representations at
levels $l$ and $m$ that satisfy the condition of having $2l$ indices
and such that $\alpha^2$ has one of the allowed values $2,0,-2,...$.
This analysis does not give information about the actual
multiplicity of the representations, which one can obtain by
comparison with listed results (see for instance the tables in ref.
\cite{Kleinschmidt:2003mf}). One can immediately see by direct
inspection that for $m=0$ one can only get a solution for $l =0$ and
$l=1$. In particular, the generator corresponding to the $l=1$
solution is an object with 2 antisymmetric indices, and the
corresponding field is the fundamental 2-form $B_2$. One has
  \begin{equation}
  F_B (l, 0) > 2 \quad {\rm for} \quad l >  2
  \end{equation}
and therefore there are no solutions for $l>2$. Also the case $l=2$
has no solution.

For $m=1$ one has solutions for $l = 0,1,...,5$, the corresponding
fields have completely antisymmetric indices and are the RR fields
$C_{2l}$ of the IIB theory. It is quite easy to see that there are
no solutions for $l \geq 6$. Therefore we have shown that the only
fields at level $m=1$ are the IIB RR forms. We will see that the
same result applies to the IIA theory. This is all in agreement with
the results of \cite{Bergshoeff:2010xc}, which show that all the RR
forms in lower dimension are obtained from dimensional reduction of
the ten-dimensional ones.

Similarly, one can consider $m=2$. In this case it is easy to see by
direct substitution that there are no solutions if $l=0$, $l=1$ and
$l=2$, as well as for $l \geq 8$. The only solutions are
  \begin{eqnarray}
  & & l = 3 \ : \ \ A_6 \nonumber \\
  & & l =4 \ : \ \ A_{7,1} \quad  A_8 \nonumber \\
  & & l = 5 \ : \ \  A_{8,2} \quad A_{9,1}  \quad 2 \times A_{10}    \nonumber\\
  & & l = 6 \ : \ \ A_{9,3} \quad A_{10,2} \nonumber \\
  & & l = 7 \ : \ \  A_{10,4} \quad , \label{m=2fieldsIIBanyl}
  \end{eqnarray}
where the multiplicity 2 of the 10-form $A_{10}$ is read from
\cite{Kleinschmidt:2003mf}. The above result summarizes all solitonic $p$-form potentials and mixed symmetry fields
contained in the IIB decomposition of $\text{E}_{11}$.

We now consider the IIA case. We can obtain the IIA theory by first
deleting node ${11}$ and then further decomposing by deleting node
${10}$, or the other way around. We denote with $l$ the level
associated with the simple root $\alpha_{11}$, and with $m$ the
level associated with the simple root $\alpha_{10}$, while the roots
from 1 to 9 are denoted by $\alpha_i$ (we deliberately use the same
notation as in the IIB case, although the decomposition of the
algebra is different). We write a generic positive root as
 \begin{equation}
 \alpha = \sum_i n_i \alpha_i + l \alpha_{11} + m \alpha_{10} \quad
 .
 \label{rootsIIA}
 \end{equation}
 We first delete node 11, and we write
  \begin{equation}
  \alpha_{11} = y - \mu_8 \label{alpha11IIAroot} \quad ,
  \end{equation}
where the vector $y$ is orthogonal to the $\text{A}_{10}$ space of
simple roots $\alpha_1,...\alpha_{10}$, and $\mu_8$ is the
fundamental weight associated to node 8 in $\text{A}_{10}$, that is
 \begin{equation}
 \mu_8 = \frac{3}{11} \left[ \alpha_1 +2 \alpha_2 + .... + 8 \alpha_8
 + \frac{16}{3} \alpha_9 + \frac{8}{3} \alpha_{10} \right]
 \end{equation}
 and
 \begin{equation}
 \mu_8^2 = \frac{24}{11} \quad .
 \end{equation}
This gives
\begin{equation}
y^2 = -\frac{2}{11} \quad . \label{ysquaredIIA}
\end{equation}
Once \eqref{alpha11IIAroot} is plugged into \eqref{rootsIIA}, one
can see that the coefficient of the root $\alpha_{10}$ is $m -
\tfrac{8}{11} l$. If we now delete node 10, we write the simple root
$\alpha_{10}$ as
 \begin{equation}
 \alpha_{10} = z - \lambda_9 \quad ,
 \end{equation}
where $z$ is a vector orthogonal to the simple roots of $\text{A}_9$
(and orthogonal to $y$ as well), and where $\lambda_9$ is the
fundamental weight of $\text{A}_9$ associated to node 9, that is
 \begin{equation}
 \lambda_9 = \frac{1}{10} \left[ \alpha_1 + 2 \alpha_2 + ... + 9 \alpha_9
 \right] \end{equation}
 and
 \begin{equation}
 \lambda_9^2 = \frac{9}{10} \quad ,
 \end{equation}
which gives
\begin{equation}
z^2 = \frac{11}{10} \quad .\label{zsquaredIIA}
\end{equation}

Repeating the same argument as for IIB, in order to have a
representation of highest weight state $\sum_i p_i \lambda_i$, we
must have
  \begin{equation}
  \alpha = \sum_i p_i \lambda_i + l y + \left(m - \frac{8}{11}l \right)
  z \quad .
  \end{equation}
Using \eqref{ysquaredIIA} and \eqref{zsquaredIIA} one gets
  \begin{equation}
  \alpha^2 = \sum_{ij} p_i A_{ij}^{-1} p_j + F_A (l,m) \quad ,
  \label{squarelegthofrootIIA}
  \end{equation}
where
  \begin{equation}
  F_A (l,m) = \frac{2}{5} l^2 + \frac{11}{10} m^2 - \frac{8}{5} ml
  \quad .
   \label{FAIIA}
  \end{equation}

One gets exactly the same result if one first deletes node 10 and
then node 11. In this case one writes
  \begin{equation}
  \alpha_{10} = \tilde{z} - \tilde{\mu}_9 \quad ,
  \label{ztildemutilde9}
  \end{equation}
where now $\tilde{z}$ is orthogonal to the space of simple roots
$\alpha_1,...,\alpha_9$ and $\alpha_{11}$ which form the algebra
$\text{D}_{10}$, and $\tilde{\mu}_9$ is the $\text{D}_{10}$
fundamental weight associated to node 9,
  \begin{equation}
  \tilde{\mu}_9 = \frac{1}{2} \left[ \alpha_1 + 2 \alpha_2 + ... + 8
  \alpha_8 + 5 \alpha_9 + 4 \alpha_{11} \right] \quad .
  \end{equation}
This gives
  \begin{equation}
  \tilde{\mu}_9^2 = \frac{5}{2} \quad ,
  \end{equation}
which implies
  \begin{equation}
  \tilde{z}^2 = - \frac{1}{2} \quad .
  \end{equation}
Once eq. \eqref{ztildemutilde9} is substituted in eq.
\eqref{alpha11IIAroot}, one can see that the coefficient in front of
the root $\alpha_{11}$ is $(l -2m)$. If we now delete node 11, this
corresponds to writing
  \begin{equation}
  \alpha_{11} = \tilde{y} - \lambda_8 \quad ,
  \end{equation}
where $\tilde{y}$ is orthogonal to the $\text{A}_9$ roots $\alpha_i$
and to $\tilde{z}$, and $\lambda_8$ is
    \begin{equation}
  \lambda_8  = \frac{1}{5} \left[ \alpha_1 + 2 \alpha_2 +3\alpha_3 +
  ...+ 8 \alpha_8 + 4 \alpha_{9} \right] \quad ,
  \end{equation}
which implies
  \begin{equation}
  \lambda_8^2 = \frac{8}{5}
  \end{equation}
and
  \begin{equation}
  \tilde{y}^2 = \frac{2}{5} \quad .
  \end{equation}
Using these results one obtains again the expression
\eqref{squarelegthofrootIIA} for the square length of the root.

We now want to analyse the representations that arise at each level.
Again, we must impose $\alpha^2 = 2,0,-2...$. We also have a
constraint on the $\text{GL}(10,\mathbb{R})$ representations coming
from imposing that the number of indices of a generator at levels
$(l,m)$ must be $2l+m$, that is
  \begin{equation}
  10 n + \sum_i p_i (10-i ) = 2 l +m \quad ,
  \end{equation}
where again $n$ is the number of groups of 10 antisymmetric indices.
For $m=0$ we get the same result as in IIB, while for $m=1$ we only
get solutions for $l=0,1,...,4$, the corresponding fields being the
RR forms, $C_{2l+1}$.

It is important to emphasise that by simply looking at
eqs.~\eqref{FBIIB} and \eqref{FAIIA} one deduces that for any given
$m$ there is a finite number of solutions, that is a finite number
of representations, in both IIA and IIB. A more careful analysis
 reveals that
  \begin{equation}
  F_B ( l + \frac{m}{2} , m ) = F_A (l , m) \quad .
  \end{equation}
This remarkable result implies that when $m$ is even the analysis of
the representations for IIB and IIA is identical, that is for any
given $m$ the representations that one gets at a given level $l$ for
IIA are the same as the ones that one gets at level $l+
\tfrac{m}{2}$ for IIB. If one considers all the representations with
a given even $m$, the IIA and IIB thus give the same result. In
particular, this implies that the $m=2$ fields are the same for both
theories, and they are given in \eqref{m=2fieldsIIBanyl}. Given that
the level must be integer, nothing can be said when $m$ is odd, and
there is no way to compare the representations of IIA with the
representations of IIB in this case.

\subsection{$\text{E}_{11}$ and T-duality}

In the previous subsection we have determined  all solitonic fields in $D=10$ dimensions, $p$-forms and mixed-symmetry fields included,
and found the following result for both IIA and IIB:
 \begin{equation}
  A_6 \qquad A_{7,1} \qquad A_8 \qquad A_{8,2} \qquad A_{9,1} \qquad
  A_{9,3} \qquad 2 \times A_{10} \qquad A_{10,2} \qquad A_{10,4} \quad .
  \label{E11alpha=-2}
  \end{equation}
We now proceed with a reduction of all these fields  on a $d$-torus, where $d=10-D$, and
we only keep the resulting $p$-forms, that is we only consider the
$D$-dimensional spacetime indices to be completely antisymmetric.
Each field, when reduced, is decomposed in forms that are
representations of $\text{SL}(d, \mathbb{R})$, and we will show that
summing for each form the representations of $\text{SL}(d,
\mathbb{R})$ resulting from all the 10-dimensional fields we obtain
representations of the T-duality group $\text{SO}(d,d)$.

We consider $D \geq 3$, while the cases $D<3$ will be considered
separately at the end of this section. The lowest form that one can
get is a $D-4$-form (this form only exists for $D \geq 4$), which
corresponds to $A_6$ with the highest possible amount of internal
indices:
  \begin{equation}
  A_{D-4, i_1 ... i_{d}} \rightarrow {\bf 1} \quad .
  \end{equation}
This is a singlet of $\text{SL}(d,\mathbb{R})$, which is of course
also a singlet of $\text{SO}(d,d)$.

The $D-3$ forms come from $A_6$,
  \begin{equation}
  A_{D-3, i_1 ... i_{d-1}} \rightarrow {\bf \overline{d}}
  \end{equation}
and from $A_{7,1}$,
   \begin{equation}
   A_{D-3, i_1 ...i_d, j} \rightarrow {\bf d} \quad .
   \end{equation}
This gives the vector representation of $\text{SO}(d,d)$ as results
from
  \begin{equation}
  {\bf 2 d} = {\bf d} \oplus {\bf \overline{d}} \quad .
  \end{equation}

Next we consider the $D-2$-forms. They come from $A_6$,
  \begin{equation}
  A_{D-2, i_1 ... i_{d-2}} \rightarrow ({\bf \overline{d} \otimes
  \overline{d}})_A \quad , \label{firstD-2form}
  \end{equation}
from $A_{7,1}$,
   \begin{equation}
   A_{D-2, i_1 ...i_{d-1}, j} \rightarrow {\bf d}\otimes {\bf
   \overline{d}} \quad , \label{secondD-2form}
   \end{equation}
from $A_8$,
  \begin{equation}
  A_{D-2  , i_1 ...i_d} \rightarrow {\bf 1}
  \end{equation}
and from $A_{8,2}$,
  \begin{equation}
  A_{D-2, i_1 i_2} \rightarrow ( {\bf d} \otimes {\bf d})_A \quad .
  \label{lastD-2form}
  \end{equation}
Here and in the rest of this section we denote with the suffix $A$
the antisymmetrised product of the representations. Summing up all
the representations of the $D-2$-forms gives the $\text{SO}(d,d)$
representations $ {\bf 1} \oplus ({\bf 2d \otimes 2d} )_A$, where
 \begin{equation}
 ({\bf 2d \otimes 2d} )_A =( {\bf d
 \otimes d })_A \oplus ( {\bf d \otimes \overline{d}} ) \oplus (
 {\bf \overline{d} \otimes \overline{d}})_A \quad .
 \end{equation}

We then consider the $D-1$-forms. From $A_6$ we get
   \begin{equation}
   A_{D-1, i_1 ... i_{d-3}} \rightarrow ({\bf \overline{d} \otimes \overline{d} \otimes \overline{d}})_A \quad ,
   \end{equation}
from $A_{7,1}$ we get
     \begin{equation}
   A_{D-1, i_1 ...i_{d-2}, j} \rightarrow {\bf d} \otimes ({\bf \overline{d}\otimes \overline{d}})_A  \quad ,
   \end{equation}
from $A_8$ we get
   \begin{equation}
   A_{D-1, i_1 ...i_{d-1}} \rightarrow {\bf \overline{d}} \quad ,
   \end{equation}
from $A_{8,2}$ we get
     \begin{equation}
   A_{D-1, i_1 ...i_{d-1}, j_1 j_2} \rightarrow {\bf \overline{d}} \otimes ({\bf d \otimes d })_A \quad ,
   \end{equation}
from $A_{9,1}$ we get
   \begin{equation}
   A_{D-1, i_1 ...i_{d}, j} \rightarrow {\bf d }  \quad ,
   \end{equation}
and finally from $A_{9,3 }$ we get
   \begin{equation}
   A_{D-1, i_1 ...i_{d}, j_1 j_2 j_3} \rightarrow ({\bf d \otimes d \otimes d})_A  \quad
   .
  \end{equation}
This gives ${\bf 2d} \oplus ({\bf 2d \otimes 2d \otimes 2d})_A $ of
$\text{SO}(d,d)$, where
  \begin{equation}
  ({\bf 2d \otimes 2d \otimes 2d})_A = ({\bf d \otimes d \otimes
  d})_A \oplus [ {\bf \overline{d}} \otimes ({\bf {d} \otimes
  {d}})_A ]\oplus [ {\bf d} \otimes ({\bf \overline{d} \otimes
  \overline{d}})_A ]\oplus ({\bf \overline{d} \otimes \overline{d} \otimes
  \overline{d}})_A \quad .
  \end{equation}

Finally, we consider the $D$-forms. All the fields in
eq.~\eqref{E11alpha=-2} contribute, and we list here the
$\text{SL}(d,\mathbb{R})$ representations for all the fields, in the
same order as they appear in eq. \eqref{E11alpha=-2}:
  \begin{eqnarray}
  & & ({\bf \overline{d} \otimes \overline{d} \otimes \overline{d}
  \otimes \overline{d}})_A \nonumber \\
  & & {\bf d} \otimes ({\bf \overline{d} \otimes \overline{d}
  \otimes \overline{d} })_A \nonumber \\
  & & ({\bf \overline{d} \otimes \overline{d} })_A \nonumber \\
  & & ({\bf d \otimes d })_A \otimes ({\bf \overline{d} \otimes
  \overline{d}})_A \nonumber \\
  & & {\bf d \otimes \overline{d}} \nonumber \\
  & & {\bf \overline{d}} \otimes ({\bf d \otimes d \otimes d })_A
  \nonumber \\
  & & 2 \times {\bf 1}\nonumber \\
  & & ({\bf d \otimes d })_A \nonumber \\
  & & ({\bf d\otimes d \otimes d \otimes d })_A \quad .
  \end{eqnarray}
This sums up to ${\bf 1} \oplus {\bf 1} \oplus ({\bf 2d \otimes 2d
})_A \oplus ({\bf 2d \otimes 2d \otimes 2d \otimes 2d })_A $ of
$\text{SO}(d,d)$. In particular, the representation corresponding to
four antisymmetrised $\text{SO}(d,d)$ indices decomposes as
  \begin{eqnarray}
  & & ({\bf 2d \otimes 2d \otimes 2d \otimes 2d})_A = ({\bf d \otimes d \otimes
  d \otimes d})_A  \oplus [ {\bf \overline{d}} \otimes ({\bf {d} \otimes
  {d} \otimes d})_A ] \nonumber \\
  & & \oplus [ ({\bf d \otimes
  d})_A \otimes ({\bf \overline{d} \otimes \overline{d}})_A ] \oplus[ {\bf d} \otimes ({\bf \overline{d} \otimes
  \overline{d} \otimes \overline{d}})_A ]
  \oplus   ({\bf \overline{d} \otimes \overline{d} \otimes
  \overline{d}\otimes \overline{d}})_A \quad .
  \end{eqnarray}

Summarizing, we find that the complete result for the solitonic $p$-form potentials precisely coincides with the one of
section 2 which is  summarised in Table \ref{NSformsanyD}. Besides,
this general analysis also shows in a very elegant and concise way
that the same set of solitonic fields occurs in four and three dimensions,
as anticipated in Table \ref{NSformsanyD}.

To complete the analysis, we also derive the forms that arise from
the reduction of the fields in eq. \eqref{E11alpha=-2} to $D<3$. We
first consider the case $D=2$. In the derivation of the scalars,
that is the 0-forms, with respect to the general derivation of the
$D-2$ forms in higher dimensions given in eqs.
\eqref{firstD-2form}-\eqref{lastD-2form}, it is eq.
\eqref{secondD-2form} that gets modified. Indeed, given that all the
indices of the field $A_{7,1}$ are internal, we have to use the fact
that the $\text{GL}(10,\mathbb{R})$ representation of the field is
irreducible, which results in the ${\bf 63}$ of
$\text{SL}(8,\mathbb{R})$ after reduction, while eq.
\eqref{secondD-2form} applied to the $D=2$ case would have given an
additional singlet. All the other representations in $D=2$ are
unchanged. To summarise, the $D$ fields in two dimensions are
  \begin{equation}
  D_{0,AB} \qquad D_{1,A} \qquad D_{1,ABC} \qquad D_{2} \qquad
  D_{2}^\prime \qquad D_{2,AB} \qquad D_{2,ABCD}
  \quad .
  \end{equation}
Repeating the same argument, for $D=1$ one finds
  \begin{equation}
  D_{0,ABC} \qquad D_{1} \qquad D_{1,AB} \qquad D_{1,ABCD} \quad .
  \end{equation}
One can formally also reduce to zero dimensions, which gives
  \begin{equation}
  D_0 \qquad D_{0,ABCD} \quad .
  \end{equation}
This last result, not surprisingly, is what one would get by
decomposing the adjoint representation of $\text{E}_{11}$ in terms
of the $\text{D}_{10}$ algebra which results from deleting node 10
in the Dynkin diagram of Fig. \ref{E11Dynkindiagram} and reading the
result at level 2.

\section{Conclusions}

In this paper we extended our previous work on the T-duality
covariant formulation of WZ terms corresponding to Fundamental
Branes and D-Branes to include the WZ terms of String Solitons, see
eq.~\eqref{WZtermuniversalexpression}. This led to two
distinguishing features which did not occur in the analysis of the
Fundamental Branes and D-branes. First of all, not all solitonic
potentials of supergravity correspond to a supersymmetric String
Soliton. This is due to the fact that the construction of a
gauge-invariant WZ term requires the introduction of a number of
worldvolume potentials that do not always fit into a worldvolume
multiplet with 16 supercharges. First, we have determined the
T-duality multiplets which contain the String Solitons, see Table
\ref{result}. Next, we identified the conjugacy classes within these
T-duality multiplets to which the String Solitons belong. These
conjugacy classes can be defined by specifying a certain set of
lightlike directions, see eq.~\eqref{informula}. Their dimensions
have been determined, see Table \ref{conjugacyclasses}.

The second distinguishing feature is that the supersymmetric String
Solitons fill out T-duality conjugacy classes, whose ten-dimensional
origin cannot be understood by considering the standard NS5A and
NS5B branes of string theory alone. The missing components can be
understood as arising from the reduction of a number of mixed
symmetry fields that are precisely the ones predicted by the very
extended Kac-Moody algebra $\text{E}_{11}$. This is only formally so
since the mixed symmetry fields, with the present technology, can
only be defined for linearised supersymmetry.

A special example of a mixed symmetry solitonic potential, predicted
by $\text{E}_{11}$, is the dual
graviton~\cite{West:2001as}.\footnote{ Another nice application of
the mixed symmetry potentials, predicted by $\text{E}_{11}$, is the
%%Eric2%%
understanding, be it at the linearized level only, of the 11-dimensional origin of the IIA 9-form RR
potential that is dual to Romans mass parameter $m$
\cite{westcommonorigin}. This is achieved by the mixed symmetry
potential $A_{10,1,1}$, with 10 antisymmetric indices and 2
symmetric ones.} In $D=10$ dimensions this is a mixed symmetry
potential  $A_{7,1}$ which has 8 indices but is only antisymmetric
in the first 7 indices. Although this dual graviton can only be
defined for linearised gravity and linearised supersymmetry
\cite{Bergshoeff:2008vc}, upon reduction to $9$ dimensions it gives
rise to a  solitonic 6-form and 7-form potential that are part of
the non-linear 9-dimensional supergravity theory. These potentials
are the duals of the Kaluza-Klein (KK) vector and KK scalar,
respectively. In particular, the 5-brane charged under the 6-form is
the reduction in the isometry direction of the KK5A (KK5B) monopole,
whose worldvolume theory is described by a vector (tensor) multiplet
\cite{Bergshoeff:1997gy}. This brane, together with the unwrapped
NS5A (NS5B) forms a vector of the nine-dimensional T-duality group
$\text{SO}(1,1)$, which splits in selfdual and anti-selfdual
representations. In lower dimensions the picture is analogous. In
particular in $D$ dimensions the KK monopole, reduced in the
isometry direction and wrapped on a $d-1$-dimensional torus, and the
NS5-brane, reduced in a transverse direction and  wrapped on a
$d-1$-dimensional torus, give $D-4$-branes that transform as vectors
of $\text{SO}(d,d)$. The important point is that, in order to obtain
the {\sl full} T-duality vector representation one needs to include
the dual graviton.\,\footnote{The dual graviton is special in the
sense that there is a corresponding KK monopole solution which can
be described in terms of the metric. Similar solutions do not exist
for the other mixed symmetry fields that are needed to fill out the
T-duality multiplets.}  The same applies to all the other
mixed-symmetry potentials predicted by $\text{E}_{11}$. They are
needed to understand the 10-dimensional origin of the different
soliton conjugacy classes.

It is interesting to compare the relation between branes and doubled
geometry in more detail. For the Fundamental Branes and the D-branes
we found that the WZ term of the Fundamental 0-branes (see
eq.~\eqref{WZ0branes}), does not contain extra scalars, i.e.~they
are insensitive for the doubled geometry. In contrast, the WZ term
of the Fundamental String (see eq.~\eqref{WZstring}) and the
D-branes (see eq.~\eqref{WZterm2}) depends on twice as many extra
scalars as there are compactified dimensions, i.e.~the Fundamental
String and the D-branes feel the doubled geometry. For the
%eric
Solitonic Branes the situation is slightly different. These branes
depend on the extra scalars via the world-volume curvature ${\cal
F}_{1,A}$ with the index $A$ {\sl uncontracted}. In section 5 we
have seen that this implies that  a particular Solitonic Brane,
which transforms as an anti-symmetric tensor with $m$ indices under
T-duality, only depends on $m$ out of the $2d$ extra scalars. The
doubled geometry, or T-duality covariance, implies that the T-dual
rotated Solitonic Brane will sense the T-dual rotated extra scalars,
but again this rotated brane will only depend on $m$ out of the $2d$
extra scalars. The fact that there is no need to impose duality
relations on the extra scalars, like in the case of the Fundamental
String \cite{Hull:2004in}, should make the construction of a
kappa-symmetric action easier.

The String Solitons, with brane tension $T\sim (g_s)^{-2}$, are just
one step in a whole family of interconnected branes with tension
$T\sim (g_s)^{\alpha}$ and $\alpha=0\,,-1\,,-2\,, \cdots$ etc.  The
next set of branes in this family, corresponding to $\alpha=-3$, are
much harder to understand. There is a crucial difference between the
fields with $\alpha \leq -3$ and the ones with  $\alpha=0\,,-1$ and
$-2$ at the level of the gauge algebra. The solitonic fields, given
in Table \ref{NSformsanyD}, have the same structure in any
dimension, and given that their transformations contain RR fields in
the form of $\text{SO}(d,d)$ spinor bilinears, the whole analysis of
the gauge algebra can be performed in a  general way which is the
same in any dimension, as was shown in section 4. For the $\alpha
=-3 $ fields the situation is different, because the requirement of
gauge invariance involves the cancellation of terms containing three
$\alpha =-1$ objects. This is achieved in each dimension using Fierz
identities of spinors of $\text{SO}(d,d)$, which implies that in
%eric
this case the analysis is dimension-dependent.

An example of an $\alpha=-3$ field is the field $E_8$ of IIB, which
is the S-dual of the RR 8-form $C_8$. Its field strength and gauge
transformations are
  \begin{eqnarray}
  & & K_9 = d E_8 + G_3 D_6- \tfrac{1}{2} F_7 C_2 \nonumber \\
 & &  \delta E_8 = d \Xi_7 + G_3 \Lambda_5 - \tfrac{1}{2} F_7
 \lambda_1 \quad ,
 \end{eqnarray}
and one can easily write down a corresponding WZ term, which
contains the world volume fields $c_1$ and $d_5$ together with two
embedding scalars. Imposing electromagnetic duality between $c_1$
and $d_5$ one obtains a vector plus two scalars, which is the
bosonic sector of a vector multiplet on an 8-dimensional world
volume, and the corresponding brane is the S-dual of the D7-brane.
As can be seen from Tables \ref{qD=9}-\ref{qD=5}, the $\alpha=-3$
field of lowest rank is always $E_{D-2,\dot{a}}$, and one can show
that this field always gives a WZ term corresponding to a
supersymmetric brane. This brane belongs to the anti-chiral spinor
$\text{SO}(d,d)$ representation, which contains the double
dimensional reduction of the S-dual of the D7-brane of IIB. One can
also perform direct dimensional reductions of this object, and
correspondingly one expects to find branes associated to $\alpha=-3
$ fields of higher rank. All these branes can be seen as the
endpoints of Fundamental Branes, D-branes and String Solitons, and
it would be interesting to determine their T-duality representations
and to investigate them in more detail.

We just saw an example of a supersymmetric $\alpha=-3$ brane which
was the S-dual of a Dirichlet brane.  By performing similar
U-duality rotations on the other branes one can easily construct
examples of supersymmetric branes with $\alpha \le -3$. The minimum
values of $\alpha$ one can obtain in this way are indicated in Table
\ref{maximum}. The cases $D=3$ and $D=4$, where the lowest values of
$\alpha$ are obtained, are a bit special. While in dimensions higher
than four the U-duality group $\text{E}_{d+1(d+1)}$ decomposes as
  \begin{equation}
  \text{E}_{d+1(d+1)} \supset \text{SO}(d,d) \times \mathbb{R}^+
  \quad ,
  \end{equation}
in four dimensions one has
  \begin{equation}
  \text{E}_{7(7)} \supset \text{SO}(6,6) \times \text{SL}(2,\mathbb{R})
  \end{equation}
and in three dimensions one has
  \begin{equation}
  \text{E}_{8(8)} \supset \text{SO}(8,8)\quad .
  \end{equation}
This implies that in order to obtain tables equivalent to Tables
\ref{qD=9}-\ref{qD=5}, and in particular in order to determine the
dilaton weight of the fields, one has to perform a further
decomposition. The U-duality representations of all the forms in
$D=4$ and $D=3$ are given in
\cite{Riccioni:2007au,Bergshoeff:2007qi}. In four dimensions the
$\text{SL}(2,\mathbb{R})$ symmetry implies that for each $p$-form,
the T-duality representation corresponding to $\alpha$ and the one
corresponding to $-\alpha -2p$ are the same, and the lowest value of
$\alpha$, corresponding to a 4-form, is $-7$, as can be deduced from
the fact that the highest value of $\alpha$ for a 4-form is $-1$. In
three dimensions, for each $p$-form the T-duality representation
corresponding to $\alpha$ and the one corresponding to $-\alpha -4p$
are conjugate, and the lowest value of $\alpha$, corresponding to a
3-form, is $-11$.

\begin{table}[t]
\begin{center}
\begin{tabular}{|c|c|c|c|c|c|c|c|c|c|}
\hline
D&IIA&IIB&9&8&7&6&5&4&3\\[.1truecm]
\hline \rule[-1mm]{0mm}{6mm} $\alpha_{\text{min}}$&$-2$ &$-4$&$-4$&$-4$&$-4$&$-5$&$-5$&$-7$&$-11$\\[.05truecm]
\hline
\end{tabular}
\end{center}
  \caption{\sl This table gives the minimum value $\alpha_{\text{min}}$ corresponding
  to the potentials
  of maximal supergravity in $D$ dimensions.}
  \label{maximum}
\end{table}

The fact that the D7-brane, with $\alpha=-1$, and the S-dual of the
D7-brane, with $\alpha=-3$, are related to each other under
U-duality, implies that the worldvolume dynamics of both branes is
described by  the same supermultiplet, which in this case is a
vector multiplet. In view of this it is convenient to  classify
branes according to the supermultiplet that governs their
worldvolume dynamics. Since we have only three different multiplets
(scalar, vector and (self-dual) tensor) we only have three different
kind of branes: Scalar, Vector and Tensor Branes. Using this
terminology, the Fundamental Branes are Scalar Branes, while the
Dirichlet brane are Vector Branes.\,\footnote{This terminology is
not unique for low-dimensional worldvolumes since for
three-dimensional worldvolumes a vector is dual to a scalar and for
two-dimensional worldvolumes a vector is dual to an integration
constant.} Among the String Solitons we have Vector and Tensor
Branes. All Vector Solitons can be obtained by a U-duality rotation
of a Vector Dirichlet Brane. Note, however, that their T-duality
representations are not the same. In this sense Vector Solitons are
not truly independent branes. This is different from the  Tensor
Solitons which stand on themselves and cannot be obtained by a
U-duality rotation of a Scalar or Vector brane.

To conclude, there are a huge number of branes in string theory. In
this paper we have concentrated on just three classes of them: the
Fundamental Branes, D-branes and Solitons. These are by far the best
understood branes. Table \ref{maximum} shows that there are many
other non-perturbative branes, whose present status is unclear. It
would be interesting to bring some order in them and to investigate
whether there is some role to play by these suggestive branes.
\bigskip

\section*{Acknowledgements}

E.B. wishes to thank King's College London and F.R. wishes to thank
the University of Groningen for hospitality. The work of F.R. was
supported by the STFC rolling grant ST/G000/395/1.

\vskip 1.5cm

\appendix

\section{Spinor representations of $\text{SO}(d,d)$}

In this appendix we summarise a few useful properties of the spinors
of $\text{SO}(d,d)$, where $d=10-D$. We will partly follow
\cite{VanProeyen:1999ni,Riccioni:2002cp}, although we will be using
a different basis for the gamma matrices. Dirac spinors of
$\text{SO}(d,d)$ have $2^d$ components. Introducing gamma matrices
$\Gamma_A\ (A=1\,\cdots,2d)$, satisfying the Clifford algebra
  \begin{equation}
  \{ \Gamma_A , \Gamma_B \} = 2 \eta_{AB}\,,
  \end{equation}
where $\eta_{AB}$ is the Minkowski metric with signature $(d,d)$,
one defines the (unitary) charge conjugation matrix $C$, whose
symmetry property is
  \begin{equation}
   C^T = -\epsilon\, C \quad , \label{Csymmetryproperty}
  \end{equation}
where $\epsilon = \pm 1$, such that
  \begin{equation}
  C \Gamma_A C^\dagger = - \eta \Gamma_A^T \quad ,
  \label{definitionofC}
  \end{equation}
where $\eta = \pm 1$. The matrix $C$ is thus symmetric if $\epsilon
= -1$ and antisymmetric if $\epsilon =1$, and the symmetry property
of  $C \Gamma_A$ is determined by the product of the two parameters
$\epsilon$ and $\eta$,
  \begin{equation}
  (C\Gamma_A)^T = \epsilon\eta\,
  (C\Gamma_A)\,.
  \end{equation}
This enables one to calculate the symmetry property of any matrix
$C\Gamma_{A_1\cdots A_n}$. Requiring that the total number of
symmetric and antisymmetric matrices equals $\tfrac{1}{2}2^d
(2^d+1)$ and $\tfrac{1}{2}2^d( 2^d-1)$, respectively, one finds, for
each value of $d$, two solutions for the pair $\epsilon\,,\eta$ such
that eqs. \eqref{Csymmetryproperty} and \eqref{definitionofC} are
satisfied. These solutions are listed in Table \ref{spinors}.
\begin{table}[t]
\begin{center}
\begin{tabular}{|l|c|c|l|c|c|}
\hline
$d$&$\epsilon$ & $\eta$& $C$ &$C\Gamma_A$&$[C,\Gamma_\star\}$\\[.1truecm]
\hline
1,5&$+1$ & $+1$& A & S & AC\\[.1truecm]
&$-1$& $-1$& S & S & \\[.1truecm]
\hline
2,6&$+1$&$-1$ &A&A&C\\[.1truecm]
&$+1$&$+1$ &A&S&\\[.1truecm]
\hline
3,7&$-1$& $+ 1$ &S&A&AC\\[.1truecm]
&$+1$& $- 1$ &A&A&\\[.1truecm]
\hline
4,8&$-1$&$-1$ &S&S&C\\[.1truecm]
&$-1$&$+1$ &S&A&\\[.1truecm]
\hline
\end{tabular}
\end{center}
\caption  {{\sl  The possible values of $\epsilon$ and $\eta$ for
various $d$. The symmetry properties of $C$ and $C\Gamma_A$ are
indicated with} S {\sl
 (symmetric) and} A {\sl  (anti-symmetric).
 The  sixth column indicates whether the charge conjugation matrix commutes} (C) {\sl
 or anti-commutes} (AC) {\sl  with $\Gamma_\star$. In each dimension, the charge conjugation
 matrix corresponding to the first line is $C_1$
 and the one corresponding to the second line is $C_2$.
 }}\label{spinors}
\end{table}
 \vskip .1truecm

Chiral spinors can be defined by introducing the matrix
$\Gamma_\star$, with $\Gamma_\star\Gamma_\star=1$, which is
proportional to the product of all other gamma-matrices:
\begin{equation}
\Gamma_\star = (-)^d\,\Gamma_1\cdots \Gamma_{2d}\,.
\label{definitionofgammastar}
\end{equation}
For all values of $D$ one can define Majorana-Weyl spinors with
$2^{d-1}$ real components. In all dimensions we are using a Weyl
basis, so that a spinor $\lambda_\alpha$ decomposes according to
\begin{equation}
\lambda_\alpha =
\begin{pmatrix}
\lambda_a\\
\lambda_{\dot a}
\end{pmatrix} \quad ,
\end{equation}
where the Dirac spinor index $\alpha$, $\alpha = 1, ..., 2^d$ splits
in the two indices $a$ and $\dot{a}$ denoting spinors of opposite
chirality, where $a,\dot{a}=1, ...,2^{d-1}$. The $\Gamma_A$ and
$\Gamma_\star$ matrices have the form
  \begin{equation}
  ( \Gamma_A )_\alpha{}^\beta=  \left( \begin{array}{cc}
  0  &  (\Gamma_A)_a{}^{\dot{b}} \\
  (\Gamma_A)_{\dot{a}}{}^b & 0 \end{array} \right) \quad ,\hskip 1truecm
  (\Gamma_\star )_\alpha{}^\beta = \begin{pmatrix}
 ( \unity_{2^{d-1}})_a{}^b&0\\
  0&-(\unity_{2^{d-1}})_{\dot a}{}^{\dot b}
  \end{pmatrix}\,. \label{chiralformofchiralmatrix}
  \end{equation}
In the Weyl basis the charge conjugation matrix is given by
\begin{equation}
d\ \text{odd}\,:\ \ C^{\alpha \beta} = \begin{pmatrix} 0&C^{a\dot
a}\cr C^{\dot a a}&0\cr
\end{pmatrix}\,,\hskip 2truecm
d\ \text{even}\,: C^{\alpha \beta} = \begin{pmatrix} C^{ab}&0\cr
0&C^{\dot a\dot b}\cr
\end{pmatrix}\,.\label{Cmatrixforminanyd}
\end{equation}
Finally, a useful identity is
 \begin{equation}
  \Gamma_{A_1\cdots A_d} = {(-)^{d}\over d!}\epsilon_{A_1\cdots
 A_{d}B_1\cdots B_{d}}\, \Gamma_*\Gamma^{B_{d}\cdots B_1}\,.
  \end{equation}

To derive all the properties of the gamma matrices and the charge
conjugation matrices that we have listed, it is convenient to work
with an explicit representation of the gamma matrices. We first
derive this representation in the Euclidean case, that is for
$\text{SO}(2d)$, and we denote the corresponding gamma matrices with
$\gamma_A$, to distinguish them from the ones of the maximally
non-compact case. If $d=1$, the Clifford algebra is simply satisfied
if $\gamma_1 = \sigma_1$ and $\gamma_2 = \sigma_2$, where
$\sigma_i$, $i=1,2,3$ are the Pauli matrices, satisfying $\sigma_i
\sigma_j = \delta_{ij} + i \epsilon_{ijk} \sigma_k$. To get the
gamma matrices for $d=2$ one considers the tensor product of Pauli
matrices and the two-dimensional identity matrix. One introduces
$\gamma_3$ and $\gamma_4$, that start as $\gamma_2$, and to make
these three matrices anticommuting one considers a tensor product
with the three different Pauli matrices, that is $\gamma_{i+1} =
\sigma_2 \otimes \sigma_i$. The Clifford algebra is then satisfied
if additionally $\gamma_1 = \sigma_1 \otimes \unity_2$. This
procedure can be induced to any $d$. Given the gamma matrices
$\gamma_A$, $A=1,...,2d$, in $2d$ dimensions, that are made of
tensor products of $d$ $2\times 2$ matrices, one considers the
matrices $\gamma_A \otimes \unity_2$, for $A = 1,...,2d-1$, together
with $\gamma_{2d} \otimes \sigma_1$, $\gamma_{2d} \otimes \sigma_2$
and $\gamma_{2d} \otimes \sigma_3$. These matrices satisfy the
Clifford algebra in $2(d+1)$ dimensions. This procedure leads to the
following result
\begin{eqnarray}
\gamma_1&=&\sigma_1\otimes \unity_2 \otimes \unity_2 \otimes
\unity_2 \otimes \unity_2 \otimes \dots
\nonumber\\
 \gamma_2&=&\sigma_2\otimes \sigma_1 \otimes \unity_2 \otimes \unity_2\otimes \unity_2 \otimes \dots
\nonumber\\
\gamma_3&=& \sigma_2\otimes \sigma_2 \otimes \unity_2
\otimes \unity_2 \otimes \unity_2 \otimes \dots\nonumber\\
\gamma_4&=&\sigma_2\otimes \sigma_3
\otimes \sigma_1 \otimes \unity_2 \otimes \unity_2 \otimes \dots\nonumber\\
\gamma_5&=&\sigma_2\otimes
\sigma_3 \otimes \sigma_2 \otimes \unity_2 \otimes \unity_2\otimes \dots\nonumber\\
\gamma_6 &=&\sigma_2\otimes \sigma_3 \otimes\sigma_3\otimes\sigma_1\otimes\unity_2\otimes \dots\label{representationClifford}\\
\gamma_7 &=&\sigma_2\otimes\sigma_3\otimes\sigma_3\otimes \sigma_2\otimes\unity_2\otimes \dots\nonumber\\
\gamma_8 &=&\sigma_2\otimes \sigma_3\otimes\sigma_3\otimes \sigma_3\otimes\sigma_1\otimes \dots\nonumber\\
\gamma_9 &=&\sigma_2\otimes\sigma_3\otimes\sigma_3\otimes\sigma_3\otimes\sigma_2 \otimes\dots\nonumber\\
\gamma_{10} &=&\sigma_2\otimes\sigma_3 \otimes\sigma_3\otimes\sigma_3\otimes\sigma_3\otimes\dots\nonumber\\
\vdots \ &=&\hskip 1.5truecm \vdots\nonumber\\
\nonumber
\end{eqnarray}
where it is understood that for ${\rm SO}(2d)$ one uses
$\gamma_1\,,\cdots,\gamma_{2d}$ and takes in each expression the
first $d$ factors of $2\times 2$ matrices.

In this basis the matrices with odd index $\gamma_{2n+1}$ are real
and symmetric, while the matrices with even index $\gamma_{2n}$ are
imaginary and antisymmetric. To go to the $(d,d)$ signature, we
define
  \begin{equation}
  \Gamma_{2n+1 } = \gamma_{2n+1} \qquad \quad \Gamma_{2n} = i
  \gamma_{2n} \label{from2dtoddmultiplyi}
  \end{equation}
which implies that all the matrices $\Gamma_n$ are real, and are
symmetric for $n$ odd and antisymmetric for $n$ even. The basis we
have chosen is a Weyl basis, and the reader can verify that defining
the chirality matrix as in \eqref{definitionofgammastar} one obtains
\begin{equation}
\Gamma_\star =
\sigma_3\otimes\unity_2\otimes\unity_2\otimes\unity_2\otimes \dots
\quad ,
\end{equation}
which is indeed as in eq.~\eqref{chiralformofchiralmatrix}.

If $d$ is even, one finds that the charge conjugation matrix
  \begin{equation}
  C_1 = \unity_2 \otimes \sigma_2 \otimes \sigma_1 \otimes \sigma_2
  \otimes \sigma_1 \otimes... \otimes \sigma_1 \otimes \sigma_2
  \end{equation}
satisfies eq.~\eqref{definitionofC} with $\eta=-1$. One can also
consider the charge conjugation matrix
  \begin{equation}
  C_2 = \sigma_3 \otimes \sigma_2 \otimes \sigma_1 \otimes \sigma_2
  \otimes \sigma_1 \otimes... \otimes \sigma_1 \otimes \sigma_2
  \quad ,
  \end{equation}
which satisfies eq. \eqref{definitionofC} with $\eta=+1$. Both these
matrices have the same symmetry properties, and they are symmetric
if $d/2$ is even and antisymmetric if $d/2$ is odd. Both these
matrices commute with $\Gamma_\star$.

If $d$ is odd, one finds that the matrix
  \begin{equation}
  C_1 = \sigma_2 \otimes \sigma_1 \otimes \sigma_2 \otimes \sigma_1
  \otimes \sigma_2 \otimes... \otimes \sigma_1 \otimes \sigma_2
  \end{equation}
satisfies eq. \eqref{definitionofC} with $\eta=+1$, and it is
symmetric if $(d+1)/2$ is even and antisymmetric if  $(d+1)/2$ is
odd. The other solution of eq. \eqref{definitionofC} for $d$ odd is
  \begin{equation}
  C_2 = \sigma_1 \otimes \sigma_1 \otimes \sigma_2 \otimes \sigma_1
  \otimes \sigma_2 \otimes... \otimes \sigma_1 \otimes \sigma_2
  \quad ,
  \end{equation}
where in this case $\eta= -1$ and the symmetry properties are
opposite to $C_1$. Both these matrices anticommute with
$\Gamma_\star$. For any $d$, the product $C_1 C_2 $ is proportional
to $\Gamma_\star$. All the properties we have derived here are
listed in Table \ref{spinors}.

The $C$ matrices we have defined are hermitian, that is they are
real if symmetric and imaginary if antisymmetric. Given that all the
gamma matrices are real one can consider spinors with real
components. When $C$ is real, given the real spinor $\lambda_\alpha$
one can then define the real spinor $\overline{\lambda}^\alpha$ from
  \begin{equation}
  \overline{\lambda}^\alpha = \lambda_\beta C^{\beta \alpha} \quad .
  \end{equation}
Similarly, when $C$ is imaginary one can define
  \begin{equation}
  \overline{\lambda}^\alpha = i \lambda_\beta C^{\beta \alpha} \quad .
  \end{equation}
We now define the conjugate of $\Gamma_A \lambda$ (without loss of
generality we can assume here that $C$ is real). One has
  \begin{equation}
  \overline{(\Gamma_A \lambda )}^\alpha = \lambda_\gamma (\Gamma_A
  )_\beta{}^\gamma C^{\beta \alpha} =  - \epsilon \lambda_\gamma (C \Gamma_A )^{\alpha \gamma} = - \eta
   ( \overline{\lambda} \Gamma_A )^\alpha \quad .
  \end{equation}
The same relation occurs when $C$ is imaginary. In this paper we
have made in all dimensions the choice for $C$ which gives $\eta
=-1$. This implies that we have always used the relation
  \begin{equation}
  \overline{(\Gamma_A \lambda )}^\alpha =
   ( \overline{\lambda} \Gamma_A )^\alpha \quad .\label{lambdabarCrelation}
  \end{equation}
This also implies that we have chosen $C$ to be imaginary for $d=2$
and $d=3$, while $C$ is real for $d=1$, $d=4$ and $d=5$, as can be
deduced by looking at the values of $\epsilon$ corresponding to
$\eta=-1$ in Table \ref{spinors}.

In this paper we have considered bilinears of $\text{SO}(d,d)$
spinors. In particular, in section 5 we have counted the number of
world-volume degrees of freedom of various brane effective actions
by looking at the world-volume fields that appear in the WZ term. In
order to perform this counting, one has to determine the number of
different components of the $\text{SO}(d,d)$ spinors that appear in
a bilinear containing $m$ Gamma matrices with fixed vector
$\text{SO}(d,d)$ indices. We now want to show that to perform this
computation one has to go to light-cone vector indices of the
T-duality group. We denote these indices by $1\pm, 2\pm ,...,d\pm$.
If $m=1$, we therefore take the single gamma matrix to be along such
directions. If $m=2$, we will see that one has to take the product
of two Gamma matrices in the directions $n_1 \pm , n_2\pm$, with
$n_1 \neq n_2$. These directions form a conjugacy class inside the
representation corresponding to two antisymmetric $\text{SO}(d,d)$
indices. Similarly, for $m=3$ and $m=4$ one has to take the the
products of Gamma matrices in the directions $n_1\pm, n_2
\pm,...,n_m \pm$, with $n_1,n_2,...,n_m$ all different. For any $m$,
the dimension of this conjugacy class is
  \begin{equation}
  2^m {d \choose m} \quad ,
  \end{equation}
and it is equal to the dimension of the corresponding
representation, ${2d \choose m}$, only for $m=0$ and $m=1$. The
dimension of the conjugacy class for $m=2$, $m=3$ and $m=4$ is given
in Table \ref{conjugacyclasses} for $d=2,...,5$.

The products of Gamma matrices in the conjugacy classes defined
above all have the property of being nilpotent. We now want to show
that spinor bilinears formed out of these matrices relate $2^{d-m}$
spinor components, with $m \leq d$, that is $2^{d-m-1}$ of one
chirality and $2^{d-m-1}$ of the opposite chirality. As we will see,
the case $m=d$ is special because a nilpotent product of $m$
light-cone Gamma matrices  maps a generic spinor to a single
component of a given chirality. The case $m>d$ is related to the
previous case by using the epsilon symbol, but we are not interested
in this case because it never gives rise to WZ terms that satisfy
our criteria.

To show that the nilpotent product of $m$ light-cone Gamma matrices,
$m\leq d$, maps a $2^d$-component spinor to $2^{d-m}$ components, we
make a specific choice of basis and we show that in this basis the
light-cone matrices have $2^{d-m}$ non-vanishing entries. For any
$d$ one can take the light-cone basis and define the nilpotent
matrices
  \begin{eqnarray}
  & & \Gamma_{1 \pm} = \frac{1}{2} \left( \Gamma_1 \pm \Gamma_{2d}
  \right) \nonumber \\
  & & \Gamma_{n\pm} = \frac{1}{2} \left( \Gamma_{2n-2} \pm
  \Gamma_{2n-1} \right) \qquad \quad n=2,...,d
  \quad , \label{lightconegammadef}
  \end{eqnarray}
where the standard Gamma matrices are those given in
\eqref{representationClifford} and \eqref{from2dtoddmultiplyi}.
Before discussing the general case, we first consider the cases
$d=1$ and $d=2$ explicitly.

For $d=1$ the Gamma matrices are $2 \times 2$ matrices, and eq.
\eqref{lightconegammadef} gives
   \begin{equation}
   \Gamma_{1+} = \begin{pmatrix}
0&1\\ 0&0
\end{pmatrix} \qquad \quad    \Gamma_{1-} = \begin{pmatrix}
0&0\\ 1&0
\end{pmatrix} \quad .
\end{equation}
One can see that, given a generic 2-component spinor, each of these
matrices maps to a given chirality. This is obvious, because these
two matrices are nothing but $\Gamma_A \pm \epsilon_{AB} \Gamma^B$.

In the $d=2$ case the matrices are $4 \times 4$, and using eq.
\eqref{lightconegammadef} one obtains
   \begin{eqnarray}
   & & \Gamma_{1+} = \begin{pmatrix}
& & 1&0\\ & & 0&0\\ 0&0& &\\ 0&1&&
\end{pmatrix} \qquad \quad    \Gamma_{1-} = \begin{pmatrix}
& & 0&0\\ & & 0&1\\ 1&0& &\\ 0&0&&
\end{pmatrix}\nonumber\\
   & & \Gamma_{2+} = \begin{pmatrix}
& & 0&0\\ & & 1&0\\ 0&0& &\\ -1&0&&
\end{pmatrix} \qquad \quad    \Gamma_{2-} = \begin{pmatrix}
& & 0&1\\ & & 0&0\\ 0&-1& &\\ 0&0&&
\end{pmatrix} \quad , \label{appendixlightconegammad=1}
\end{eqnarray}
where here and in the following the blocks that are left blank have
all zero entries. One can see that each of these matrices maps a
generic 4-component spinor to one component of each chirality. We
next consider $m=2$, that is the antisymmetric product of two Gamma
matrices. Given the 6 independent $\Gamma_{AB}$ matrices, one can
choose in light-cone coordinates the basis $\Gamma_{1\pm} \Gamma_{2
\pm}$, $(\Gamma_{1+} \Gamma_{1-} -\Gamma_{1-} \Gamma_{1+} )$ and
$(\Gamma_{2+} \Gamma_{2-} - \Gamma_{2-} \Gamma_{2+})$, where only
the first four matrices belong to the nilpotent conjugacy class. In
particular, one has
 \begin{equation}
 \Gamma_{1+} \Gamma_{2+} = \begin{pmatrix}
 0&0& & \\ 0 & 0&  &\\ && 0&0\\ & & 1& 0\end{pmatrix}
 \quad ,
 \end{equation}
which has only one non-zero entry, and thus maps a spinor to one
component of a given chirality. The same applies to the other three
matrices obtained by taking the other combinations of pluses and
minuses, while the reader can verify that this does not apply to
(linear combinations of) the other two matrices $(\Gamma_{1+}
\Gamma_{1-} -\Gamma_{1-} \Gamma_{1+} )$ and $(\Gamma_{2+}
\Gamma_{2-} - \Gamma_{2-} \Gamma_{2+})$.

We now briefly consider the general case $d>2$. We explicitly write
the matrices $\Gamma_{2\pm}$ and $\Gamma_{3\pm}$. For
$\Gamma_{2\pm}$ using \eqref{representationClifford} and
\eqref{from2dtoddmultiplyi} one gets
  \begin{equation}
  \Gamma_{2+} =
\begin{pmatrix}
 & & $\textbb{0}$ & $\textbb{0}$ \\
 & & \unity & $\textbb{0}$  \\
$\textbb{0}$ &$\textbb{0}$ &&\\
 -\unity &$\textbb{0}$ & &
\end{pmatrix} \quad , \qquad \Gamma_{2-} =
\begin{pmatrix}
 & & $\textbb{0}$ & \unity\\
 & & $\textbb{0}$  & $\textbb{0}$  \\
$\textbb{0}$ & - \unity  &&\\
$\textbb{0}$ &$\textbb{0}$ &&
\end{pmatrix} \label{appendixlightconegammad>22+2-}
 \quad ,
\end{equation}
where \textbb{0} and $\unity$ are the $2^{d-2} \times 2^{d-2}$ zero
and identity matrices. Similarly, for $\Gamma_{3\pm}$ one gets
  \begin{equation}
  \Gamma_{3+} =
\begin{pmatrix}
 & & & & $\textbb{0}$ & $\textbb{0}$ & & \\
 & & &&\unity & $\textbb{0}$ && \\
& & & & & & $\textbb{0}$ &$\textbb{0}$ \\
& & & & & & -\unity & $\textbb{0}$ \\
$\textbb{0}$ & $\textbb{0}$& & & & & & \\
-\unity & $\textbb{0}$ & & & & & & \\
& & $\textbb{0}$ & $\textbb{0}$& & & & \\
& & \unity & $\textbb{0}$ & & &
\end{pmatrix} \label{appendixlightconegammad>23+}
\end{equation}
and
  \begin{equation}
  \Gamma_{3-} =
\begin{pmatrix}
 & & & & $\textbb{0}$ & \unity & & \\
 & & && $\textbb{0}$ & $\textbb{0}$ && \\
& & & & & & $\textbb{0}$ & -\unity \\
& & & & & & $\textbb{0}$ & $\textbb{0}$ \\
$\textbb{0}$ & -\unity & & & & & & \\
$\textbb{0}$ & $\textbb{0}$ & & & & & & \\
& & $\textbb{0}$ & \unity & & & & \\
& & $\textbb{0}$ & $\textbb{0}$ & & &
\end{pmatrix} \label{appendixlightconegammad>23-}
\end{equation}
where here \textbb{0} and $\unity$ are the $2^{d-3} \times 2^{d-3}$
zero and identity matrices. One can see that each of these matrices
halves the spinor components of each chirality. The product of the
two Gamma matrices  $\Gamma_{2+}$ and $\Gamma_{3+}$  is
  \begin{equation}
  \Gamma_{2+} \Gamma_{3+} =
\begin{pmatrix}
 & &  $\textbb{0}$ & $\textbb{0}$ & & & & \\
 & &  $\textbb{0}$ & $\textbb{0}$ & & & & \\
$\textbb{0}$ &$\textbb{0}$ & & & & & & \\
 -\unity & $\textbb{0}$ & & & & &  &\\
& & & & & & $\textbb{0}$ & $\textbb{0}$ \\
& & & & & & $\textbb{0}$ & $\textbb{0}$ \\
& & & & $\textbb{0}$ & $\textbb{0}$& & & \\
& & & & -\unity & $\textbb{0}$ & &
\end{pmatrix} \quad , \label{gamma2+gamma3+}
\end{equation}
where again \textbb{0} and $\unity$ are the $2^{d-3} \times 2^{d-3}$
zero and identity matrices and thus this matrix maps a
$2^d$-component spinor to $2^{d-3}$ components for each chirality.
The reader can verify that the same applies to the other three
combinations of pluses and minuses, while it does not apply to
$(\Gamma_{2+} \Gamma_{2-} -  \Gamma_{2-} \Gamma_{2+})$, as well as
to $(\Gamma_{3+} \Gamma_{3-} -  \Gamma_{3-} \Gamma_{3+})$.

We consider explicitly the case $d=3$, for which the entries in eq.
\eqref{gamma2+gamma3+} are numbers. We then compute $\Gamma_{1+}$
from eq. \eqref{lightconegammadef}, that is
  \begin{equation}
  \Gamma_{1+} =
\begin{pmatrix}
 & & & & 1& 0 & & \\
 & & && 0 & 0 && \\
& & & & & & 0 & 0 \\
& & & & & & 0 & 1 \\
0 & 0 & & & & & & \\
0 & 1 & & & & & & \\
& & 1 & 0 & & & & \\
& & 0 & 0 & & &
\end{pmatrix} \quad ,
\end{equation}
and we finally consider the product of the three light-cone Gamma
matrices
  \begin{equation}
  \Gamma_{1+} \Gamma_{2+} \Gamma_{3+}=
\begin{pmatrix}
 & & & & &  & 0&0 \\
 & & &&  &  & 0 & 0\\
& & & & 0&0 &  &  \\
& & & & -1&0 &  &  \\
&  &0 &0 & & & & \\
 &  &0 &0 & & & & \\
0&0 &  &  & & & & \\
0& 0&  &  & & &
\end{pmatrix} \quad ,
\end{equation}
which as expected has only one non-zero entry, and thus maps an
8-component spinor to a single component of a given chirality.

A careful analysis of the Gamma matrices defined in eqs.
\eqref{representationClifford} and \eqref{from2dtoddmultiplyi} and
the light-cone Gamma matrices of eq. \eqref{lightconegammadef}
should convince the reader that the result is completely general and
applies to all cases.

\end{document}